\documentclass[aps,prx,twocolumn,amssymb,amsfonts,superscriptaddress]{revtex4-2}  

\usepackage{natbib}
\usepackage{graphicx}
\usepackage{amsmath}
\usepackage[inline]{enumitem}
\usepackage{textcomp} 
\usepackage[mathscr]{euscript}
\usepackage{bbm}
\usepackage{setspace} 
\usepackage{xcolor}

\usepackage{xr}

\def\in{_\text{in}}
\def\out{_\text{out}}

\def\m{_\text{m}}

\def\rout{\mathbf{r\out}}
\def\rin{\mathbf{r\in}}

\def\rm{\mathbf{r\m}}

\def\tin{\theta\in}

\def\Rrr{\mathbf{R}_{\mathbf{r}\mathbf{r}}}
\def\Rxx{\mathbf{R}_{xx}}
\def\Rrrbar{\overline{\mathbf{R}}_{\mathbf{r}\mathbf{r}}}

\newcommand*{\la}[1]{\textcolor{black}{#1}}
\newcommand*{\alex}[1]{\textcolor{black}{#1}}

\newcommand*{\willy}[1]{\textcolor{black}{#1}}
\newcommand*{\al}[1]{\textcolor{black}{#1}}
\newcommand*{\mat}[1]{\textcolor{black}{#1}}
\newcommand*{\laura}[1]{\textcolor{black}{#1}}
\newcommand*{\rev}[1]{\textcolor{black}{#1}}
\newcommand*{\just}[1]{\textcolor{black}{#1}}
\newcommand*{\lc}[1]{\textcolor{black}{#1}}
\newcommand*{\revv}[1]{\textcolor{black}{#1}}

\begin{document}

\title{Ultrasound Matrix Imaging—Part I: The focused reflection matrix, the $F$-factor and the role of multiple scattering.}

\author{William Lambert}
\affiliation{Institut Langevin, ESPCI Paris, CNRS, PSL University, 1 rue Jussieu, 75005 Paris, France}
	\affiliation{Hologic / SuperSonic Imagine, 135 Rue Emilien Gautier, 13290 Aix-en-Provence, France}
\author{Justine Robin}
\affiliation{Institut Langevin, ESPCI Paris, CNRS, PSL University, 1 rue Jussieu, 75005 Paris, France}
\affiliation{Physics for Medicine Paris, INSERM, CNRS, ESPCI Paris, PSL University, 17 rue Moreau, 75012 Paris, France}
\author{Laura A. Cobus}
\affiliation{Institut Langevin, ESPCI Paris, CNRS, PSL University, 1 rue Jussieu, 75005 Paris, France}
\affiliation{Dodd-Walls Centre for Photonic and Quantum Technologies and Department of Physics, University of Auckland, Private Bag 92019, Auckland 1010, New Zealand}
\author{Mathias Fink}
\affiliation{Institut Langevin, ESPCI Paris, CNRS, PSL University, 1 rue Jussieu, 75005 Paris, France}
\author{{Alexandre Aubry}}
\email{alexandre.aubry@espci.fr}
\affiliation{Institut Langevin, ESPCI Paris, CNRS, PSL University, 1 rue Jussieu, 75005 Paris, France}





\begin{abstract}
This is the first article in a series of two dealing with a matrix approach  {for} aberration quantification and correction in ultrasound imaging. Advanced synthetic beamforming relies on a double focusing operation at transmission and reception on each point of the medium. Ultrasound matrix imaging (UMI) consists in decoupling the location of these transmitted and received focal spots. The response between those virtual transducers form the so-called focused reflection matrix that actually contains much more information than a confocal ultrasound image. In this paper, a time-frequency analysis of this matrix is performed, which highlights the single and multiple scattering contributions as well as the impact of aberrations in the monochromatic and broadband regimes. Interestingly, this analysis enables the measurement of the incoherent input-output point spread function at any pixel of this image. \rev{A fitting process enables the quantification of the single scattering, multiple scattering and noise components in the image. From the single scattering contribution, a focusing criterion is defined\la{, and its evolution used to quantify the amount of aberration throughout the ultrasound image. }
In contrast} to the state-of-the-art coherence factor, this new indicator is robust to multiple scattering and electronic noise, thereby providing a contrasted map of the focusing quality at a much better transverse resolution. After a validation of the proof-of-concept based on time-domain simulations, UMI is applied to the \textit{in-vivo} study of a human calf. \revv{Beyond this specific example, UMI opens a new route for speed-of-sound and scattering quantification in ultrasound imaging.}
\end{abstract}

\maketitle

\section{Introduction}
\label{sec:introduction}
To investigate soft tissues \alex{in ultrasound imaging}, a sequence of \alex{incident} waves is used to insonify the medium. 
\la{Inside the medium, the waves encounter short-scale fluctuations of acoustic impedance, generating}
back-scattered \alex{echoes} that are used to build an ultrasound image. Conventionally, this estimation of the medium reflectivity is performed using the process of delay-and-sum (DAS) beamforming, which relies on a \rev{coherent summation of the signals associated with each echo generated by scatterers in the medium.} 
\just{Signals from a particular echo are} selected by computing the time-of-flight associated \alex{with} the forward and return travel paths of the ultrasonic wave between \rev{each transducer} and the image voxel. From a physical point of view, time delays in transmission are used to concentrate the ultrasound wave on a focal area \alex{whose size is ideally only limited} by diffraction. Time delays at reception select echoes coming from this excited area. This process falls into the so-called confocal imaging techniques, meaning that, for each point of the image, a double focusing operation is performed.

The 
\la{critical} step of computing the time-of-flight for each insonification and each focal point is achieved in any clinical device by assuming the medium as homogeneous with a constant speed of sound. This assumption is 
\la{necessary in order to achieve the rapidity required for real-time imaging; however,} it may not be valid for some configurations \la{in which} long-scale fluctuations of the medium \la{speed of sound} impact wave propagation \cite{hinkelman1997measurements}. In soft tissues, \la{such fluctuations are around $5$\%, as the speed of sound typically ranges from 1400~m/s (\textit{e.g.} fat tissues) to 1650~m/s (\textit{e.g.} skin, muscle tissues)~\cite{duck1990acoustic}.
In such 
situations, the incident focal spot spreads beyond the diffraction-limited area, the exciting pressure field at this focusing point is reduced, and undesired echoes are generated by surrounding areas. In} reception, 
\rev{the application of 
\lc{an incorrect} time delay profile mixes echoes which originate from neighbor\lc{ing} points in the medium, resulting in a distorted point spread function (PSF) at the output.} These aberrating effects can strongly degrade the image resolution and contrast. For highly heterogeneous media, \la{such} aberrations may impact the diagnosis of the medical exam or limit the capability to image some organs. \la{A \laura{classic} example of this \laura{effect} 
is liver imaging of difficult-to-image patients. Because the ultrasonic waves must travel through successive layers of skin, fat, and muscle tissue before reaching the liver,} 
both the incident and reflected wave-fronts undergo strong aberrations~\cite{hinkelman1998effect}.

To assess the quality of focus and monitor the convergence of aberration correction techniques, Mallart and Fink \cite{Mallart1994} \alex{introduced a focusing parameter $C$, known as the coherence factor in the ultrasound community~\cite{Hollman}. Based on the van Cittert-Zernike theorem,} this indicator \alex{is linked to} the spatial coherence of the back-scattered field measured by the probe for a focused transmit beam. 
This method assumes 
\la{that images are dominated by ultrasonic speckle -- a grainy, noise-like texture which is} generated by a medium of random reflectivity. This \la{assumption is in fact often valid} 
in medical ultrasound, \la{where} soft tissues are \la{composed} of randomly distributed scatterers that are unresolved at ultraso\la{nic} 
frequencies. In that case, each transmitted focusing wave excites an ensemble of scatterers \la{which are} randomly distributed within the focal spot. The back-scattered wavefront results from the random superposition of \la{the} echoes generated by each of those scatterers. 
\la{If the focus is perfect, the time delays used to focus (in transmission and in reception) are exactly the same as the actual round-trip time-of-flight of the waves which arrive at, and are backscattered by, scatterers located within the resolution cell.} 
The spatial coherence of the received signal is then maximal\just{;} 
\la{the associated coherence factor} $C$ tends towards 
$2/3$ in the speckle regime \rev{and in \lc{the} absence of clutter} \lc{(contributions from multiple scattering to the detected backscattered signal)}. 
\la{A decrease in the quality of focus is matched by a decrease in $C$; thus, this indicator is popular in the literature for the evaluation of ultrasonic focusing quality. }

\alex{Very recently, \al{ultrasound matrix imaging (UMI)} has been proposed for a quantitative mapping of aberrations in ultrasound imaging~\cite{lambert2020reflection}. Experimentally, 
\la{the first step in this approach is to record} the reflection matrix associated with the \just{imaged} medium. \willy{This matrix contains \al{the response of the medium recorded by each transducer of the probe, for a set of illuminations}}. 
\willy{Depending on the problem \la{one is facing}, this matrix can be investigated in \alex{different} bases}. Here, for imaging purpose\la{s}, the reflection matrix will be projected in\la{to} a focused basis. In contrast with conventional ultrasound imaging that relies on 
\la{confocal} beamforming, the idea \la{here is to} 
\la{apply independent focused beamforming procedures at the input and output of the reflection matrix. This process}  
yields a focused reflection (FR) matrix that contains the responses between virtual transducers synthesized \just{directly in the medium} from the transmitted and received focal spots. \la{The FR matrix} holds much more information on the medium than a conventional ultrasound image. Importantly, \rev{it can be leveraged 
\lc{to discriminate between} the single scattering, multiple scattering and noise contributions. From the 
\lc{former,} a new focusing criterion $F$ is introduced to assess the local focusing quality, notably in the speckle regime, for any pixel of the image.} While this indicator shows \alex{some} similarities with the coherence factor introduced by Mallart and Fink \cite{Mallart1994}, \al{the parameter $F$ constitutes a much more sensitive and \rev{spatially-resolved probe} of the focusing quality}. 
In the quest for local correction of distributed aberrations~\cite{ali2018distributed,Jaeger2015SosMap,Rau2019,bendjador2020svd,lambert2020distortion}, this parameter $F$ can be of particular interest as it can play the role of a guide star for any pixel of the ultrasound image. More generally, the FR matrix is a key operator for imaging applications~\cite{Badon2016,Badon2019,Blondel2018,lambert2020distortion,Touma2020}. The accompanying article~\cite{lambert_ieee} will present in details a matrix method of aberration correction in which the FR matrix plays a pivotal role.}

The current paper describes the mathematical construction and physical meaning of the FR matrix. \la{In particular, it is shown how the FR matrix can be exploited to probe the local ultrasound focusing quality, independently of the medium reflectivity.} 
\la{The FR matrix was originally introduced in Ref.~\cite{lambert2020reflection}; here, it is investigated in more detail, and its properties are demonstrated \rev{by means of time domain simulations and applied to} \textit{in-vivo} ultrasound imaging of the calf of a human subject.} 
First, a time-frequency analysis of the FR matrix is performed\la{, and its} 
mathematical expression is derived rigorously 
\la{for} the monochromatic and broadband regimes. \rev{In 
\lc{the broadband regime}, the analytical expression of the FR matrix goes beyond the result of Ref.~\cite{lambert2020reflection} that consisted in a simple expansion of the monochromatic result at the central frequency}. \lc{In this article, we discuss the} 
relative weight of single and multiple scattering contributions \lc{and} the impact of aberrations 
\la{for} 
each regime. While Ref.~\cite{lambert2020reflection} only considered virtual transducers belonging to the same transverse plane, 
\la{here we} investigate the impulse responses between focusing points located at any depth [see Fig.~\ref{fig:2Dcommon}(a)]. A local incoherent input-output PSF \la{can \just{thus} be extracted from} 
the FR matrix 
\la{along both} the axial and transverse directions. \rev{Going beyond Ref.~\cite{lambert2020reflection}, a novel fitting process of this PSF is also proposed to discriminate \lc{between} the single and multiple scattering components of the reflected wave-field. 
\lc{This process} leads to a new definition of the focusing factor $F$ 
\just{as a scaling factor in the transverse direction
of the input-output PSF compared to its diffraction-limited (ideal) counterpart.} 
\lc{(Note that in the latter quantity, the local frequency spectrum of ultrasonic data is taken into account).} 
\lc{This entire} process is first validated by means of numerical simulations. 
\just{The} 
comparison with the state-of-the-art $C$-map shows a drastic gain in terms of contrast and resolution for the local evaluation of the focusing quality. \rev{The gain in contrast} is due to the fact that, unlike the coherence factor $C$, $F$ is robust to both multiple scattering and electronic noise. \just{The gain in transverse resolution is provided by the input-output focusing operation inherent to UMI while the $C-$factor is estimated from a single focusing operation. }
\lc{We then \just{illustrate} this property of $F$ by applying} UMI to in-vivo ultrasound data. 
\just{A} predominant incoherent background \just{due to multiple scattering and electronic noise} yields a very weak and badly contrasted coherence factor $C$\just{, while} the focusing factor $F$ highlights different regions 
\lc{in which the focus is almost perfect ($F\sim 1$), and others in which} the focusing  process is degraded by irregular layers of muscle, fat and skin at shallow depth\lc{s}.}

The paper is structured as follows: Section~\ref{sec1} \rev{describes} the experimental procedure to acquire the reflection matrix \rev{and the numerical simulations that will be used in the paper}. 
In Sec.~\ref{sec2}, the FR matrix is \rev{investigated} in the monochromatic regime. A 2D local input-output PSF is extracted \la{from} 
the FR matrix. The manifestation of aberrations, out-of-focus echoes and multiple scattering in this monochromatic PSF is discussed \rev{by means of numerical simulations and experimental in-vivo data}. In Sec.~\ref{sec4}, a coherent sum of the monochromatic FR matrices leads to a \al{broadband} FR matrix. 
This operation, which is equivalent to a time-gating process, suppresses contributions from out-of-focus reflectors. \rev{
\lc{In addition, as shown by numerical simulations, 
this summing process} enables a clear discrimination between the single and multiple scattering components of the reflected wave-field.} In \rev{Sec.~\ref{sec5}}, a time-frequency analysis of the FR matrix is performed and shows, in particular, the impact of absorption and scattering on the \rev{experimental} FR matrix as a function of depth and frequency. In \rev{Sec.~\ref{sec6}}, a fitting procedure is proposed to extract the relative weights of single scattering, multiple scattering and electronic noise. From the single scattering component, a focusing parameter $F$ is obtained by \rev{rescaling} the \rev{transverse evolution of the broadband input-output PSF with the ideal, diffraction-limited one. The overall fitting procedure is validated by means of the time domain simulations and applied to the in-vivo ultrasound imaging of a human calf.} 
Finally\rev{, Sec.~\ref{sec7}} 
\la{outlines} the advantages of 
$F$ compared to the coherence factor $C$ 
\la{widely} used in the literature~\cite{Mallart1994,Hollman}. Section~\ref{sec8} presents conclusions and general perspectives. 

\section{\label{sec1}Methods}
\subsection{Experimental Procedure}

\la{\al{UMI} begins with an experimental recording of the reflection matrix, $\mathbf{R}$. In principle, this measurement can be achieved using any 
type of illumination (element-by-element ~\cite{Aubry2009}, \mat{focused beams}~\cite{Aubry2008}, \textit{etc.}); 
here, a plane-wave acquisition sequence~\cite{Montaldo2009} has been arbitrarily chosen. The probe was placed in direct contact with the calf of a healthy \lc{human} volunteer, orthogonally to the muscular fibers. (This study is in conformation with the declaration of Helsinki).} 
The acquisition was performed using a medical ultrafast ultrasound scanner (Aixplorer Mach-30, Supersonic Imagine, Aix-en-Provence, France) driving a \rev{$5-18$} MHz linear transducer array containing $192$ transducers with a pitch $p=0.2$ \rev{mm} (SL18-5, Supersonic Imagine). 
\la{The 
acquisition sequence consisted of transmission of} 
$101$ steering angles spanning from $-25^{o}$ to $25^{o}$, 
\la{calculated for the hypothesis of a tissue speed of sound of} $c_0=1580$ m/s~\cite{duck1990acoustic}. The pulse repetition frequency \la{was} set at $1000$ Hz. The emitted signal \la{was} a sinusoidal burst 
\la{lasting} for three half periods of the central frequency $f_c=7.5$~MHz. For each excitation, the back-scattered signal \la{was} recorded by the $192$ transducers of the probe over a time length $\Delta t = 80$ $\mu$s \la{at} 
sampling frequency $f_s=40$ MHz. Acquired in this way, the reflection matrix is denoted\ $\mathbf{R}_{\mat{u}\theta}(t)\equiv R(\mat{u}\out,\tin,t)$, where $u\out$ defines the \mat{transverse} position of the \rev{receiving} transducer, $\tin$\al{, the incident angle of the emitted plane wave and $t$,} the time-of-flight. In the following, the subscripts \emph{in} and \emph{out} will denote the \al{transmitting and receiving} steps \mat{of the reflection matrix recording process}.
\begin{figure}
    \centering
    \includegraphics[width=1\columnwidth]{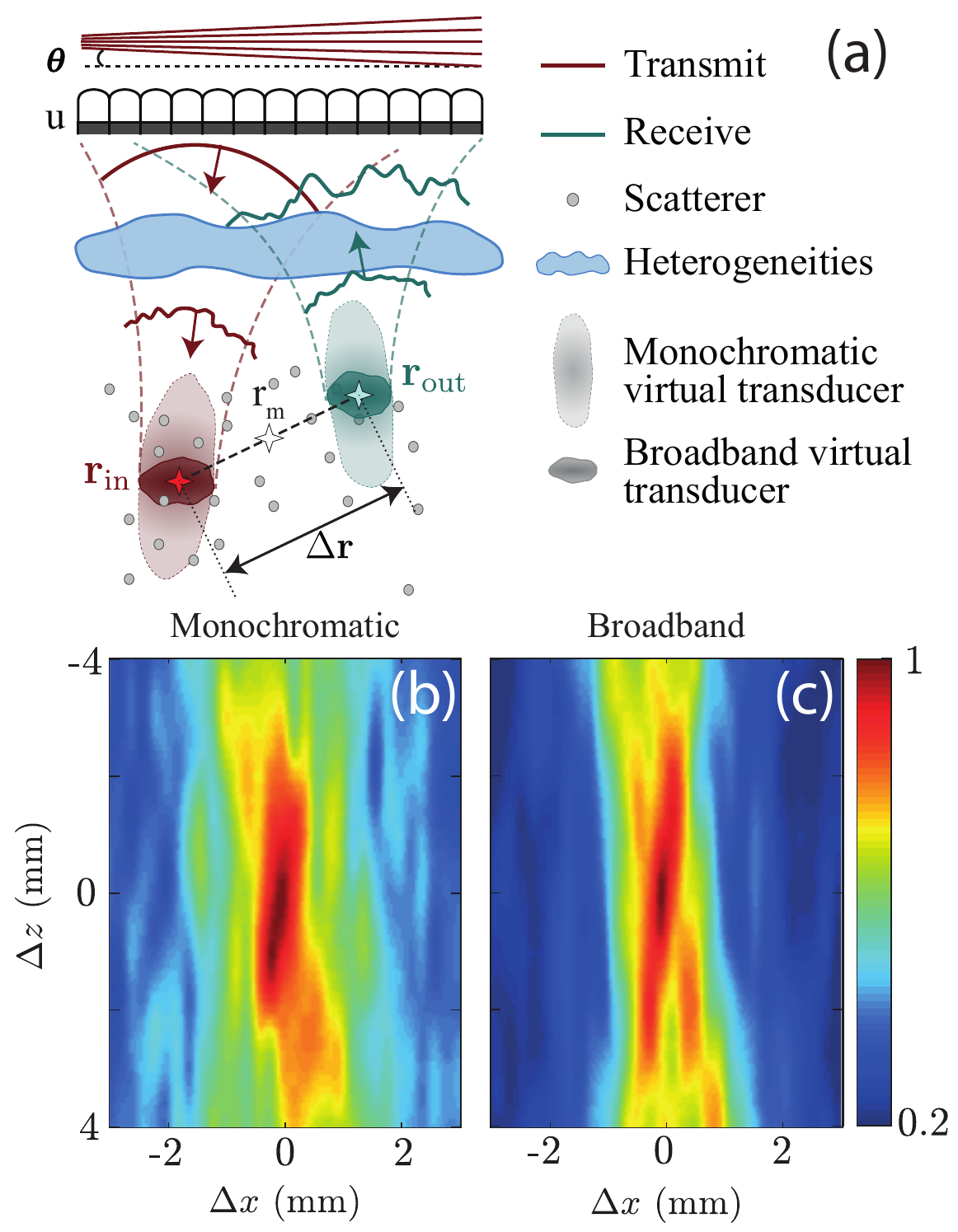}
    \caption{Principle of the \mat{FR} matrix. (a) \al{UMI} consists in splitting the location of the transmitted ($\rin$) and received focusing ($\rout$) points both in the axial and transverse directions, thereby synthesizing virtual transducers that can act as 
    source and detector\la{, respectively} at any point in the medium. In a monochromatic regime, the synthesized virtual transducer displays an elongated shape in the $z-$direction because of diffraction. In the broadband domain, the axial resolution is inversely proportional to the signal bandwidth, 
    \la{giving} a much thinner virtual transducer in the $z-$direction. (b)-(c) Monochromatic (\mat{$f=7.5$ MHz}) and broadband common mid-point intensity profile \eqref{IrDr_fromRrDr} averaged over a set of common mid-points $\mathbf{r}$ \mat{contained in the white rectangle displayed in Fig.~\ref{fig:focusing2}(a$_1$) and (a$_2$)}. Both profiles have been normalized by their maximum. }
    \label{fig:2Dcommon}
\end{figure}

\subsection{Numerical Simulations}

\begin{figure}
    \centering
    \includegraphics[width=1\columnwidth]{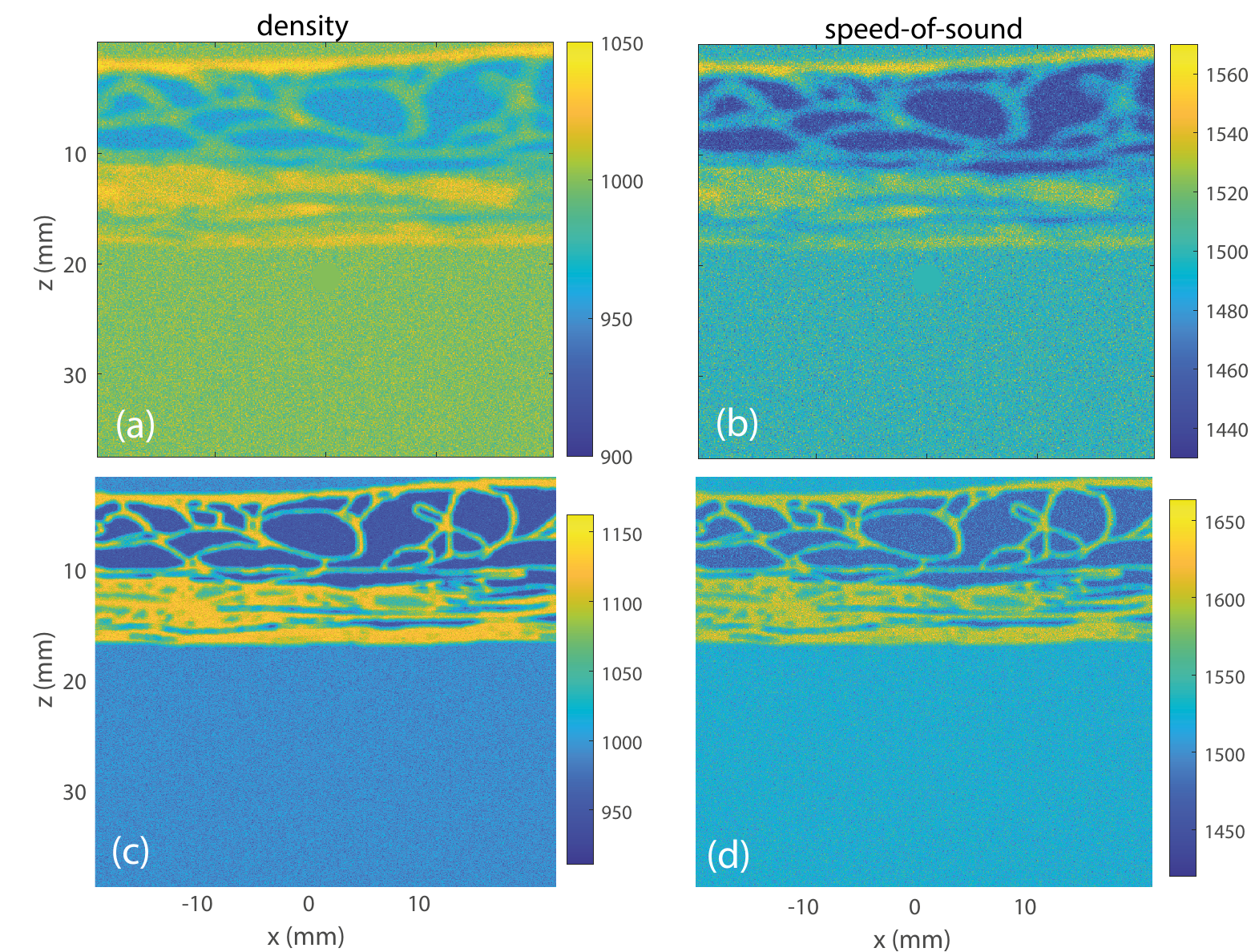}
    \caption{\revv{Acoustic properties of low speckle media simulated with k-wave. (a,b) Density and speed-of sound distributions resulting from the low-pass filtering of original maps (c,d) introduced by Mast \textit{et al.}~\cite{Mast1997}. The speckle statistics is provided in Tab.~\ref{tab:parameters}.}}
    \label{fig:rhoc}
\end{figure}

\rev{To validate our aproach, a multistatic synthetic aperture dataset has also been computed with k-Wave~\cite{Treeby2010}, a time domain simulation software based on the k-space pseudospectral method. The density and speed-of-sound distributions of the simulated medium [Figs.~\ref{fig:rhoc}(a) and (b)] mimic the aberrations undergone by ultrasound through the abdominal wall~\revv{\cite{Ali2022}. They result from a spatial low-pass filtering of the numerical tissue layers introduced by Mast \textit{et al.}~\cite{Mast1997} [Figs.~\ref{fig:rhoc}(c) and (d)] on which speckle has been superimposed. Those s}hort-scale \rev{fluctuations of these mechanical properties generate a random} wave-field characteristic of ultrasound imaging in soft tissues. A \just{2.6} mm-diameter anechoic disk is also included \rev{at a depth of $20$ mm to highlight the impact of aberrations and multiple scattering on the imaging process. To investigate the effect of multiple scattering, two k-wave simulations have been performed to mimic ultrasound imaging in 
\lc{low} and high speckle scattering regimes. The speckle strength is controlled by means of the standard deviation of the speed-of-sound and of the density, $\sigma_c$ and $\sigma_\rho$, respectively. The values considered in our simulations are reported in \just{Tab.~\ref{tab:parameters}}. In the Rayleigh regime, the impact of density fluctuations is \rev{negligible compared to speed-of-sound variations~\cite{Baydoun2015}. Under a scalar, Gaussian and bi-dimensional model of disorder, the scattering mean free path $\ell_s$ can be estimated through the following relation~\cite{Aubry2008_pdh}: $\ell_s \sim 2 (\sigma_c/c_0)^{-2} k_0^{-3} \ell_c^{-2}$, with $k_0$ the wave number and $\ell_c$ the coherence length of speckle. The latter parameter is here roughly equal to the spatial sampling used for the simulation. It yields an order of magnitude for $\ell_s$ that is reported in \just{Tab.~\ref{tab:parameters}} for each simulation.}}}

\rev{Each k-Wave simulation is performed over a two-dimensional grid. All simulation parameters such as the \rev{transducer configuration, transmit pulse, computational grid, and sampling are described in \just{Tab.~\ref{tab:parameters}}. The recording of the reflection matrix is performed as described above for the experiment. The acquisition sequence consisted of transmission of $101$ steering angles spanning from $-25^{o}$ to $25^{o}$. The emitted signal was \just{a 2.25 MHz two-cycle sinusoidal burst}. For each excitation, the back-scattered signal {was} recorded by the \just{155} transducers of the probe {at} sampling frequency $f_s=80$ MHz.}}
\begin{table}
    \renewcommand{\arraystretch}{1.4}
    \centering
            \caption{\label{tab:parameters} \just{Simulation parameters}.}
    \begin{tabular}{|c|c|c|}
     \hline
        Parameter & Low speckle & High speckle \\
        \hline
        Grid Size & 
  \multicolumn{2}{|c|}{$563 \times 620$ pixels} \\
        \hline
        Spatial sampling   & 
  \multicolumn{2}{|c|}{66.7 $\mu$m}\\
        \hline
  Time sampling   & 
  \multicolumn{2}{|c|}{12.5 ns}\\
     \hline
  Recording length  & 
  \multicolumn{2}{|c|}{63 $\mu$s}\\
     \hline
     
  Inter-element pitch   & \multicolumn{2}{|c|}{267 $\mu$m}\\
     \hline
  Emission signal amplitude  & 
  \multicolumn{2}{|c|}{0.2 MPa}\\
     \hline
  Speed-of-sound  & 
  \multicolumn{2}{|c|}{1400 -- 1600 m.s$^{-1}$}\\
  \hline
  $\sigma_c$ (m.s$^{-1}$) & 15 & 45 \\
       \hline
  Density  & 
  \multicolumn{2}{|c|}{900 -- 1070 kg.m$^{-3}$}\\
  \hline
 $ \sigma_\rho$ (kg.m$^{-3}$) & 10 & 30 \\
  \hline
   $\ell_s$ (mm) &   600 & 70 \\
     \hline
  Absorption coefficient  & 
  \multicolumn{2}{|c|}{0.15 -- 0.4 dB.MHz$^{-1.5}$.cm$^{-1}$}\\
     \hline
    \end{tabular}
\end{table}

\section{\label{sec2} Monochromatic focused reflection matrix}

\alex{The \rev{first post-processing step of UMI} consists in projecting the reflection matrix in\la{to} a focused basis. \la{This step is greatly simplified by performing beamforming operations in the frequency domain, i.e.} applying appropriate phase shifts to \la{all} frequency component\la{s} of the received signals in order to realign them at each \la{focal} point.}  A matrix formalism is particularly suitable for this operation\la{, because, in 
the frequency domain, the projection of} data from the plane-wave or transducer bases to any focal plane can be achieved with a simple matrix product \rev{previously described in Ref.~\cite{lambert2020reflection}}. \rev{Note that the focused basis corresponds to a set of vectors that are not linearly independent
\lc{; this is therefore not,} strictly speaking, a basis in a mathematical sense. 
\lc{In contrast,} the incident plane waves used to record the reflection matrix are linearly independent, but they do not form a complete basis. Nevertheless, for convenience, we will refer to these two sets of wave-fronts as focused and plane wave bases in the following.}

The result of this beamforming process is a set of focused reflection matrices $\Rrr\al{(f)}$ obtained at each frequency $f$ of the bandwidth. In contrast to standard synthetic ultrasound imaging, in which input and output focusing points coincide, the approach presented here decouples these points. In emission, the incident energy is concentrated at the focusing point $\rin$\la{; this point can thus} be seen as a virtual source\mat{~\cite{Passmann1996,Bae2000}}. Similarly, in reception, a virtual sensor is synthesized \rev{by selecting the part of the back-scattered wave-field \just{originating from the vicinity of} point $\rout$.} Therefore, each coefficient $R(\rout, \rin, f)$ of $\Rrr\al{(f)}$ contains the monochromatic responses of the medium between a set of virtual transducers. \la{The position of each virtual transducer maps onto a pixel of the ultrasound image.} {Fig.~\ref{fig:2Dcommon}\mat{(a)}} illustrates \la{this} matrix focusing process. \la{Note that, for clarity}, the input focusing operation \la{in {Fig.~\ref{fig:2Dcommon}}{(a)}} is represented by \la{a} cylindrical wave-front instead of a combination of plane waves. This is justified by the fact that \mat{a} plane wave synthetic beamforming numerically mimics a focused excitation \cite{Montaldo2009}. \rev{Note that the concept of virtual transducers is merely didactic and that, of course, they do not act as real source or sink of energy. Moreover, they are strongly directive: in the downward direction for the virtual source, in the upward direction for the virtual receiver. At last, in the monochromatic regime, the coupling between the transmit and receive beams can also occur above and below the focal depth due to spurious echos from out-of-focus scatterers.}

\begin{figure*}
    \centering
    \includegraphics[width=1\textwidth]{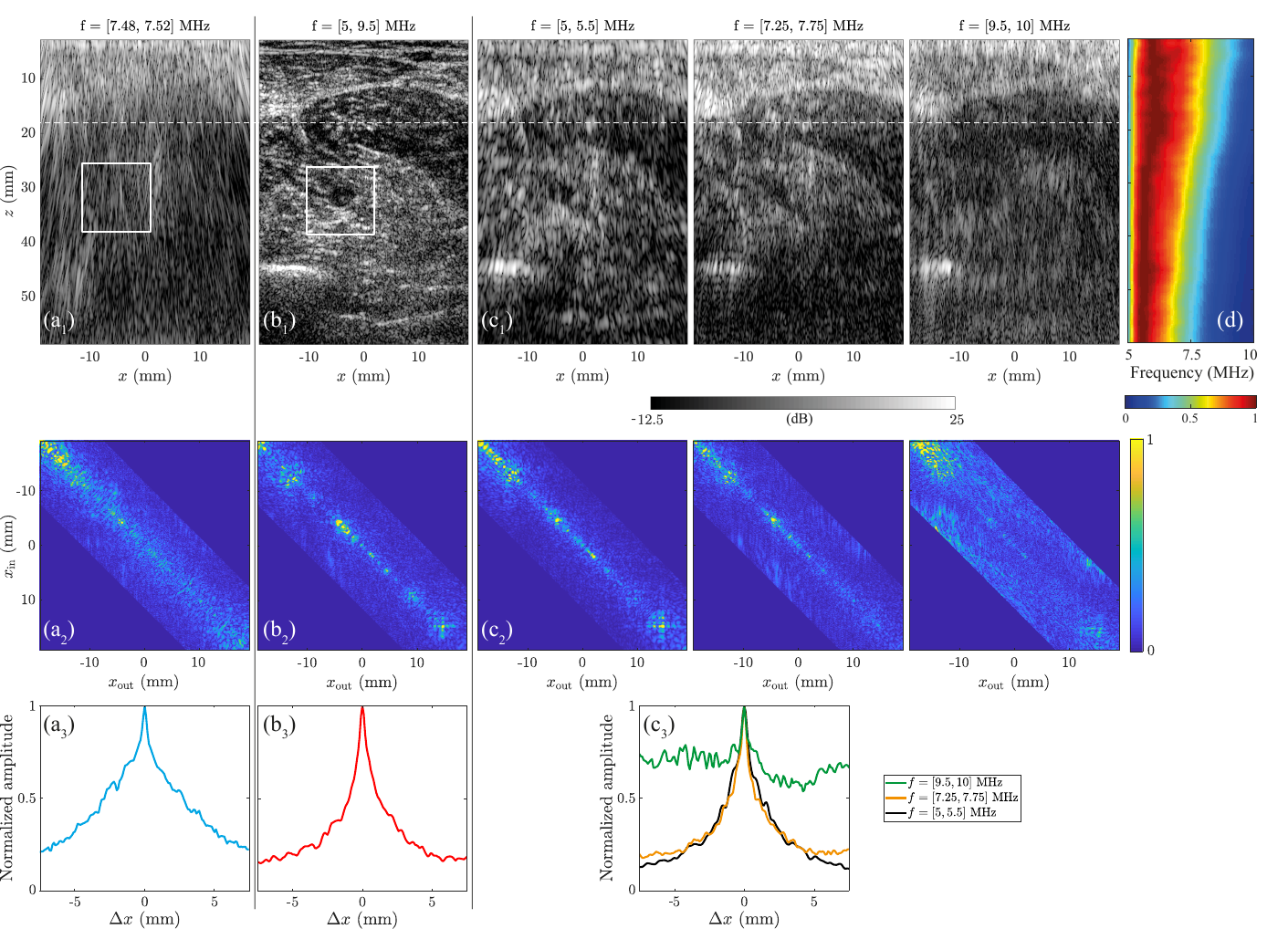}
    \caption{\alex{\al{UMI} in a monochromatic, broadband and multi-frequency regime\rev{: Experimental result}. (a) Monochromatic regime ($f=$7.5 MHz): (a$_1$) Multi-focus image (dB-scale), (a$_2$) reflection matrix $\Rxx(z)$ at $z=18$ mm \mat{[see white dashed line in (a)]}, and (a$_3$) common mid-point intensity profile $I(z,\Delta x)$ (\eqref{IrDr_fromRrDr}) averaged at the same depth. (b) Broadband regime (5-10 MHz): (b$_1$) Multi-focus image (dB-scale), (b$_2$) Broadband reflection matrix $\Rxx(z)$ at $z=18$ mm, and (b$_3$) common mid-point intensity profile $I(\mathbf{r},\Delta x)$ averaged at the same depth.
    (c) Multi frequency regime (5-5.5 MHz, 7.25-7.75 MHz, 9.5-10 MHz, from left to right): (c$_1$) Multi-focus images (dB-scale), (c$_2$) multi-frequency reflection matrices $\Rxx(z)$ at $z=18$ mm, and (c$_3$) common mid-point intensity profiles $I(\mathbf{r},\Delta x)$ averaged at the same depth. (d) Spectrogram of the confocal signals versus depth $z$ extracted from the diagonal elements of $\Rxx(z)$ (linear scale). The confocal spectrum is normalized by its maximum at each depth. \mat{The white rectangle in (a$_1$) and (a$_2$) accounts for the position of CMPs considered for the computation of the CMP intensity profiles displayed in Fig.~\ref{fig:2Dcommon}(b) and (c), respectively.}}
    }
    \label{fig:focusing2}
\end{figure*}

Fig.~\ref{fig:focusing2}(a$_2$) shows the $x$-projection $\Rxx(z,f)$ of $\Rrr(f)$ at \la{depth} $z=18$ mm and frequency $f=6$ MHz in the experiment.
$\Rxx(z,f)=[R(x\out, x\in,f,z)]$ contains the responses between virtual transducers, $\rin=(x\in,z)$ and $\rout=(x\out,z)$, located at the same depth.
Note that \rev{$\Delta x$ has to be limited} to avoid the spatial aliasing induced by the incompleteness of the plane wave illumination basis; \rev{its maximal value} $\Delta x_{\mathrm{max}}$ is inversely proportional to the angular step $\delta \theta$ of the plane wave illumination basis: $\Delta x_{\mathrm{max}}\sim \lambda_{\mathrm{max}}/(2\delta \theta)$, with $\lambda$ the wavelength. Thus, the coefficients $R(x\out, x\in,f,z)$ associated with a transverse distance $|x\out - x\in|$ larger than a superior bound $\Delta x_{\mathrm{max}}$ are not displayed.

\begin{figure*}
    \centering
    \includegraphics[width=\textwidth]{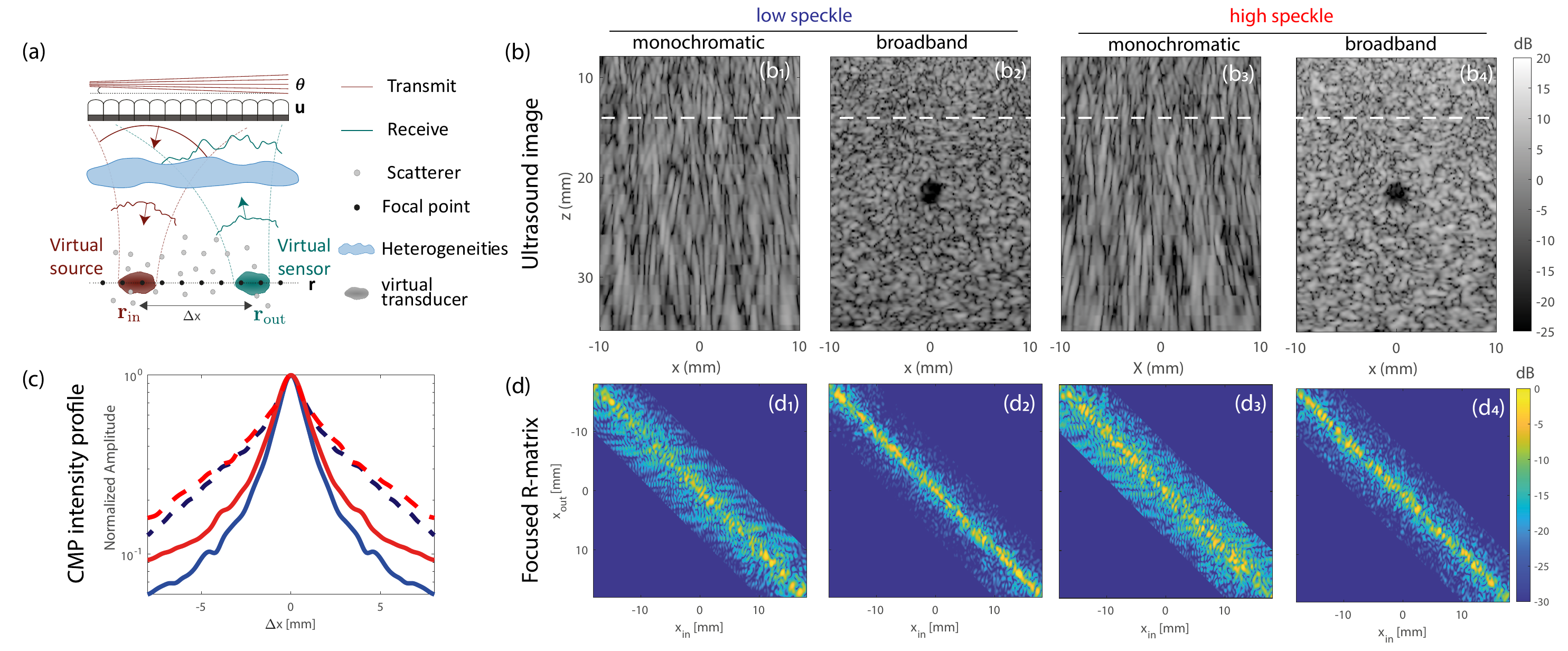}
    \caption{\rev{{UMI} in low and high speckle scattering regimes: Numerical results. (a) $x-$cross section of the {FR} matrix: The transmitted ($\rin$) and received focusing ($\rout$) points are at the same depth. (b) Multi-focus images (dB-scale) in the low (b$_1$, b$_2$) and high (b$_3$, b$_4$) speckle regimes built over a restricted (2.245-2.255 MHz) and broad (0.5-4 MHz) bandwidth, respectively. (c) Correspond CMP intensity profile $I(z,\Delta x)$ (\eqref{IrDr_fromRrDr}) at $z=13$ mm [see white dashed line in (b)] in the monochromatic (dashed lines) and broadband (continuous lines) regimes for a weak (blue lines) and high (red lines) speckle. (d) FR matrix $\Rxx(z)$ at the same depth in the low (d$_1$, d$_2$) and high (d$_3$, d$_4$) speckle regimes built over a restricted (2.245-2.255 MHz) and broad (0.5-4 MHz) bandwidth, respectively.}
    }
    \label{fig:focusing1}
\end{figure*}

\rev{Fig.~\ref{fig:focusing2}(a$_2$) shows} that most of the signal in $\Rxx(f)$ \rev{tends to concentrate} around its diagonal. 
\al{This feature is characteristic of a predominant single scattering contribution~\cite{lambert2020reflection}}. Indeed, singly-scattered echoes \rev{mainly} originate from a virtual detector $\rout$ 
\la{which is located near to} virtual source $\rin$. \rev{This is confirmed by Fig.~\ref{fig:focusing1} that compares two monochromatic FR matrices $\Rxx(f)$ ($z=13$ mm, $f=2.25$ MHz) obtained numerically in the low and high speckle scattering regimes [Fig.~\ref{fig:focusing1}(d$_1$) and (d$_3$), respectively]. While the multiple scattering rate is supposed to differ in each case, the FR matrices display a similar feature, which corroborates that single scattering is here predominant.} 

The diagonal elements of~$\Rrr(f)$, which obey $\rin=\rout$, \la{directly} \al{provide the image \mat{$\mathcal{I}(\mathbf{r},f)$}} which would be obtained via multi-focus \la{(a.k.a. confocal)} imaging at frequency $f$:
\begin{equation}
\label{imcalc}
\mathcal{I}\left(\mathbf{r},f\right)\equiv \left|R\left(\mathbf{r},\mathbf{r},f \right)\right|^2.
\end{equation}
\laura{An} example of such an image is displayed in Fig.~\ref{fig:focusing2}(a$_1$). Compared to a standard ultrasound image built from broadband signals \la{(Fig.\ \ref{fig:focusing2}b$_1$)}, this \mat{monochromatic} image displays 
\la{poor} axial resolution \mat{and is therefore difficult to exploit}. \rev{This observation is confirmed numerically whether it be in the low (Fig.~\ref{fig:focusing1}(b$_1$)) or high (Fig.~\ref{fig:focusing1}(b$_3$) speckle regime.} 
%
However, \lc{as will be shown} \mat{in the following,} \la{the off-diagonal elements of $\Rrr$ 
provide valuable} information on the physical properties of the medium, as well as on the wave focusing quality.
\\

\just{Under the Born approximation and its matrix formalism\cite{Brutt2022},} \la{$\Rrr(f)$ can be expressed theoretically} as follows\rev{~\cite{lambert2020reflection}}:
\begin{equation}
    \Rrr(f)  = \mathbf{H}^{\top}\out (f)  \times \mathbf{\Gamma}(f) \times \mathbf{H_{in}} (f),
    \label{eq:RrrMatrix__Hin_times_Hout}
\end{equation}
\rev{where the matrix $\mathbf{\Gamma}$ describes the scattering process in the \rev{sample}; in the single scattering regime, $\mathbf{\Gamma}$ is diagonal and its elements correspond to the medium reflectivity $\gamma(\mathbf{r},f)$ at frequency $f$. $\mathbf{H_\text{in}}(f)$ and $\mathbf{H_\text{out}}(f)$ are the input and output focusing matrices, respectively\rev{~\cite{lambert2020reflection}}.} Each column of $\mathbf{H}\in(f)=[H\in(\mathbf{r},\rin, f)]$ corresponds to one \mat{focused} illumination, and contains the monochromatic PSF at emission -- the spatial amplitude distribution of the input focal spot resulting from that particular illumination. Similarly, each column of $\mathbf{H}\out(f)=[H\out(\mathbf{r},\rout, f)]$ contains the spatial amplitude distribution of an output focal spot. As will be shown in the following, the spatial extension of these focal spots is a direct manifestation of the mismatch between the actual velocity distribution in the medium \rev{and the spatially-invariant velocity model used in the matrix beamforming process}. For lighter notation in the rest of this section, the frequency dependence $f$ of each physical quantity is made implicit.

Equation~\eqref{eq:RrrMatrix__Hin_times_Hout} can be rewritten in terms of matrix coefficients as follows:
\begin{equation}
\label{Rrr_intermsofh_eq}
R(\rout,\rin)=\int d \mathbf{r}  ~ H_\text{out}(\mathbf{r},\mathbf{r}_\text{out})  ~ \gamma(\mathbf{r}) ~ H_\text{in}(\mathbf{r},\mathbf{r}_\text{in}) .
\end{equation}
\rev{This equation shows that each pixel of the \just{ultrasound image (diagonal elements of $\mathbf{R_{rr}}$)}} results from a convolution between the sample reflectivity $\gamma(\mathbf{r})$ and an imaging PSF, $H(\mathbf{r}, \mathbf{r}')$, 
\la{which is itself a product of} the input and output PSFs: $H(\mathbf{r}, \mathbf{r}')=H_\text{in}(\mathbf{r}, \mathbf{r}') \times H_\text{out}(\mathbf{r}, \mathbf{r}')$.  
As ${H}\in(\mathbf{r},\mathbf{r}\in)$ and  ${H}\out(\mathbf{r},\mathbf{r}\out)$ define the characteristic size of \rev{the input focal spot} at $\rin$ and \rev{the output focal spot}  at $\rout$, 
\lc{these matrices also define} the resolution of the \lc{resulting} confocal image. In the absence of aberration, the transverse and axial dimensions of these focal spots, $\delta x_0(\mathbf{r})$ and $\delta z_0(\mathbf{r})$, are only limited by diffraction~\cite{Born}:
\begin{equation}
	\delta  x_0(\mathbf{r})=\frac{\lambda}{2 \sin[\beta(\mathbf{r})]}  \, \text{,    } \delta z_0(\mathbf{r}) = \frac{ 2\lambda }{ \sin^2[\beta(\mathbf{r})]},
	\label{eq:thResolutionMonochrom}
\end{equation}
with $\beta(\mathbf{r})$ the maximum angle 
by which each focal point is illuminated or seen by the array of transducers. \la{In ultrasound imaging, the radiation pattern of the transducers usually limits the numerical aperture to $\rev{\sim} 25^\circ$. The \rev{focal spots} thus typically display} a characteristic elongated shape in the $z-$direction \rev{[$\delta x_0<<\delta z_0$, see Fig.~\ref{fig:focusing1}(a)]}, which accounts for the \la{poor} axial resolution exhibited by the monochromatic image\rev{s} \rev{obtained experimentally [Fig.~\ref{fig:focusing2}(a$_1$)] and numerically [Figs.~\ref{fig:focusing1}(b$_1$,b$_3$)].} 

\section{Common mid-point intensity}
The off-diagonal points in $\Rrr$ can be exploited for 
\la{a quantification of the} focusing quality at any pixel of the ultrasound image. To that aim, the relevant observable is the intensity profile along each anti-diagonal of $\Rrr$~\cite{lambert2020reflection}: 
\begin{equation}
\label{IrDr_fromRrDr}
I(\rm,\Delta \mathbf{r})= \left | R(\rm +\Delta \mathbf{r} /2,\rm-\Delta \mathbf{r} /2) \right |^2.
\end{equation}
All 
\la{pairs} of points on a given anti-diagonal have the same midpoint $\rm = (\rout + \rin)/2$ , but varying spacing $\Delta \mathbf{r}=(\rout-\rin)$. In the following, $I(\al{\rm},\Delta \mathbf{r})$ is thus referred to as the common-midpoint (CMP) intensity profile. \la{To express this quantity theoretically, we first make an \rev{isoplanatic} approximation in the vicinity of each CMP $\rm$. This means that waves which focus in this region are assumed to have travelled through approximately the same areas of the medium, thereby undergoing identical phase distortions} 
\cite{lambert2020distortion,lambert_ieee}. \la{The input and/or output PSFs can then be considered to be spatially invariant within this local region.}
\willy{\al{Mathematically, 
\la{this} means that, in the vicinity of each \la{common} mid-point $\rm$, the spatial distribution of the input 
\la{or} output PSFs, $H_\text{in/out}(\mathbf{r},\mathbf{r}_\text{in/out})$,} only depend\la{s} \al{on} the relative distance between the point $\mathbf{r}$ and the focusing point $\mathbf{r}_{\text{in/out}}$. \rev{This leads us to define a local spatially-invariant PSF $  H^{(l)}_\text{in/out}$ around each common mid-point $\rm$ such that:}}
\begin{equation}
\label{isoplanatic}
H_\text{in/out}(\mathbf{r},\mathbf{r}_\text{in/out}) =  \rev{ H^{(l)}_\text{in/out}(\mathbf {r}-\mathbf{r}_\text{in/out},\rm).}
\end{equation}
\rev{The range of validity of this approximation is given by the size of each isoplanatic patch. In our numerical simulation, azimuthal and axial isoplanatic patch sizes are of the order of 3~mm and 5~mm, respectively, at 80\% correlation. \lc{For the experimental data (acquired on a human calf), we obtain similar values for isoplanatic patch size -- these were extracted from the aberration phase laws given by}  
UMI~\cite{lambert2020distortion}}

\la{We next make the assumption that, as is often the case} in ultrasound imaging, scattering is 
due to a random distribution of unresolved scatterers. Such a speckle scattering regime can be modelled by a random reflectivity:
\begin{equation}
\label{random}
\langle \gamma (\mathbf{r_1}) \gamma^*(\mathbf{r_2})\rangle = \langle \left | \gamma  \right|^2 \rangle \delta(\mathbf{r_2}-\mathbf{r_1}),
\end{equation}
where $\langle...\rangle$ denotes an ensemble average and $\delta$ is the Dirac distribution. \rev{By injecting \eqref{Rrr_intermsofh_eq} into \eqref{IrDr_fromRrDr}, then reducing the implicit double integral over $\mathbf{r}$ by the $\delta$-function in \eqref{random}, and \rev{finally using \eqref{isoplanatic} and the change of variables ($\mathbf{r} - \rm \rightarrow \mathbf{r}$), the following expression can be found for the \al{CMP} intensity:}}
\begin{multline}
  I(\rm,\Delta \mathbf{r})=  \int d \mathbf{r} ~ |\rev{H^{(l)}_\text{out}}\left (\mathbf{r}-{\Delta \mathbf{r}}/{2},\rm \right) |^2 ~ \\ |\rev{H^{(l)}_\text{in}}\left (\mathbf{r}+{\Delta \mathbf{r}}/{2},\rm \right)|^2 ~ |\gamma(\mathbf{r}+\rm)|^2.
  \label{IrDr_fromRrDr2}
\end{multline}
To smooth the intensity fluctuations due to the random reflectivity, a spatial average over a few resolution cells is required while keeping a satisfactory spatial resolution. To do so, a \rev{normalized} spatially averaged intensity profile $I_\textrm{av}(\al{\rm},\Delta \mathbf{r})$ is computed in the vicinity of each point $\al{\rm}$, such that
\begin{equation}
\label{average}
     I_\textrm{av}(\rm,\Delta \mathbf{r}) = \frac{\left \langle W_{L} (\mathbf{r}-\rm) {I}(\mathbf{r},\Delta \mathbf{r})  \right\rangle_{\mathbf{r}} }{\rev{\left \langle W_{L} (\mathbf{r}-\rm) {I}(\mathbf{r},\mathbf{0})  \right\rangle_{\mathbf{r}}}}
\end{equation}
where the symbol $\langle ... \rangle_{\mathbf{r}}$ denotes the spatial average and $W_{L}(\mathbf{r})$ is a spatial window function, such that
\begin{equation}
\label{eq:window}
W_{L}(\mathbf{r}) =\left\{
\begin{array}{ll}
1 & \, \mbox{for }\rev{\lVert\mathbf{r}\lVert}<L/2 \\
0 & \,  \mbox{otherwise}.
\end{array} 
\right. 
\end{equation}
This spatial averaging process leads to 
\lc{the replacement of} $|\gamma(\mathbf{r})|^2$ in the last equation by its ensemble average $\langle |\gamma|^2\rangle$. $I_\textrm{av}(\al{\rm},\Delta \mathbf{r} )$ then directly provides the convolution between the incoherent input and output PSFs, $|\rev{H^{(l)}_\text{in}}|^2$ and $|\rev{H^{(l)}_\text{out}|^2}$:
\begin{equation}
\label{IrDr_fromRrDr2b}
I_\textrm{av}(\rm,\Delta \mathbf{r}) \propto  \left [ |\rev{H^{(l)}_\text{in}|^2} \stackrel{\al{\Delta \mathbf{r}}}{\circledast}  |\rev{H^{(l)}_\text{out}|^2} \right] (\Delta \mathbf{r}, \rm)\lc{,}
\end{equation}
\lc{where the symbol $\stackrel{\Delta \mathbf{r}}{\circledast}$ denotes a spatial convolution over $\Delta \mathbf{r}$}.
Note that this formula only holds in the speckle regime; for a specular reflector, the CMP intensity profile is equivalent to the intensity of the coherent input-output PSF, $|\rev{H^{(l)}_\text{in}}  \stackrel{\al{\Delta}\mathbf{r}}{\circledast} \rev{H^{(l)}_\text{out}}|^2 $~\cite{lambert2020reflection}. \willy{
\la{In either case, this quantity gives two interesting pieces of information. Firstly,} \rev{the CMP intensity at $\Delta \mathbf{r} = \mathbf{0}$ \rev{is proportional to the confocal energy at focus}}. Secondly, the spatial extension of the CMP 
is linked to the lateral and axial dimensions of the input and output PSFs.} 

Fig.~\ref{fig:2Dcommon}\mat{(b)} displays an example of a \al{two dimensional} CMP intensity profile. The corresponding cross-section of this profile at $\Delta z=0$ is displayed in Fig.~\ref{fig:focusing2}(a$_3$). 
\la{This cross-section} has been averaged over a set of \al{CMPs} contained in the white rectangle of \mat{Fig.~\ref{fig:focusing2}(a$_1$)}. Similarly to the input/output PSFs $\rev{H^{(l)}_\text{in}}$ and $\rev{H^{(l)}_\text{out}}$ \eqref{eq:thResolutionMonochrom}, the incoherent input-output PSF $ |\rev{H^{(l)}_\text{in}}|^2  \willy{\stackrel{\al{\Delta}\mathbf{r}}{\circledast}} |\rev{H^{(l)}_\text{out}}|^2$ displays a cigar-like shape. However, while its axial FWHM 
is close to the diffraction limit (
$\delta z_0 \sim 2.2 $ mm, with 
\la{$\beta=25^\circ$}), its transverse FWHM \rev{($\sim 1$ mm)} is far from being ideal ($\delta x_0 \sim 0.25$ mm). 

Ideally, the CMP intensity profile would be given by
\begin{equation}
\label{I0}
I_0 (\Delta x , \rm)  =|\rev{H^{(l)}_0}|^2 \stackrel{\al{\Delta x}}{\circledast}  |\rev{H^{(l)}_0}|^2 (\Delta x , \rm),
\end{equation}    
\rev{
\lc{where $\rev{H^{(l)}_0}$ is} the diffraction-limited PSF, such that
\begin{equation}
\rev{H^{(l)}_0}(\Delta x , \rm) =\text{sinc} \left (2\pi \Delta x \sin \left [\beta(\rm) \right] /\lambda \right)
\end{equation}    
with }$\text{sinc}(x) = \sin(x)/x$. 
\rev{The CMP intensity profile} should \rev{thus} ideally display a transverse full width at half maximum (FWHM) roughly equal to \al{the diffraction-limited resolution} $\delta x_0$.

\la{The poor \rev{transverse} resolution \rev{highlighted by Fig.~\ref{fig:2Dcommon}\mat{(b)}} can be explained by several potential effects.} First, long-scale variations of the speed-of-sound can \la{give }
rise to aberrations that distort the input and output PSFs~\cite{lambert2020distortion}. Second, 
\la{if there are spurious echo\rev{e}s from out-of-focus scatterers (located above or below the focal plane), then the expansion of the resulting input\rev{/output} beams will be greater than those originating exactly at the focal plane [Fig.~\ref{fig:focusing1}(a)]. 
Note also that short-scale heterogeneities} may induce multiple scattering events that give rise to an incoherent background in the \al{CMP} intensity profile~\cite{lambert2020reflection}. 
\lc{As we will see,} the prevalence of multiple scattering events can be estimated from the amount of signal at off-diagonal elements -- an incoherent background which is a combination of multiple scattering contributions and\rev{, in the experiment,} of electronic noise. 

\rev{All these contributions} constitute a problem for imaging; in the monochromatic regime under examination here, \rev{it is} extremely difficult to discriminate between the effects of aberration, multiple scattering, and singly-scattered echoes taking place out-of-focus. In the next section, we show that the contributions from 
out-of-focus echos can be greatly reduced via a time-gating operation. This \rev{process} \rev{enables 
\lc{the recovery of \just{the standard axial resolution exhibited by} ultrasound images.}} 

\section{\label{sec4}Broadband reflection matrix}

\alex{Under the matrix formalism, 
\la{time gating} can be performed by building a broadband FR matrix $\Rrrbar$. \la{In the following, we show that b}esides improving the axial resolution and contrast of the ultrasound image, 
\la{$\Rrrbar$} 
\la{allows} a clear distinction between the \la{contributions from} single and multiple scattering.}

\alex{\alex{In} the frequency domain, the FR matrix is built by dephasing each RF signal 
\lc{such that} scattering paths \rev{whose first and last scattering events take place at $\rin$ and $\rout$ constructively interfere}. A coherent sum over the overall bandwidth $\Delta f$ can then be performed to build a broadband FR matrix:}
\begin{equation}
  \alex{  \Rrrbar (\Delta f) =  \willy{\frac{1}{\Delta f}} \int^{f_{+}}_{f_{-}} df ~ \Rrr(f)}
    \label{eq:sumRrr_freq}
\end{equation}
with $f_{\pm}=f_c \pm \Delta f/2$ and $f_c$ the central frequency of the RF signal bandwidth. 
One row of the broadband FR matrix corresponds to the situation 
\la{in which} the transmitted waves are focused 
at $\rin$\la{, creating a virtual source} while the virtual detector probes the spatial spreading of this virtual source at the \al{expected ballistic} time, \al{\textit{i.e} at time $t=0$ in the focused basis} [Fig.~\ref{fig:2Dcommon}(a)]. \la{Thus, the sum of monochromatic FR matrices over the entire bandwidth can be interpreted as a time-gating operation in which echos originating from a certain range of times-of-flight are extracted.}

\la{With the time-gating applied \eqref{eq:sumRrr_freq}, the axial resolution of the virtual transducers should be drastically improved [see Fig.~\ref{fig:2Dcommon}(a)]. To prove this assertion, 
we now derive an expression for the broadband FR matrix within the formalism of this work.  
For sake of simplicity and analytical tractability, 
paraxial and isoplanatic \la{\eqref{isoplanatic} approximations are made. 
\la{The monochromatic PSFs can be decomposed} as follows:}
\begin{equation}
\label{paraxial}
    H_\textrm{in/out}(\mathbf{r},\mathbf{r'},f)=\rev{\overline{H}_\textrm{in/out}}(\mathbf{r}-\mathbf{r'}\al{,\rm},f)~e^{j2\pi f (z-z') /c}
\end{equation}
where $\rev{\overline{H}_\textrm{in/out}}$ represents the envelope of the PSF. Injecting \eqref{Rrr_intermsofh_eq} and \eqref{paraxial} into \eqref{eq:sumRrr_freq} leads to the following expression for the coefficients of $\Rrr(\Delta f)$}:
\begin{multline}
\overline{R}(\rout,\rin,\Delta f) =  \int d \mathbf{r} ~ e^{j2\pi f_c (2z - z_\textrm{in}-z_\textrm{out})/c} \\
\text{sinc} \left [ \frac{\pi \Delta f}{c}(2z - z_\textrm{in}-z_\textrm{out})\right ] \\
\rev{\overline{H}_\textrm{out}}(\mathbf{r}-\mathbf{r}_\text{out}\al{,\rm}) ~ \gamma(\mathbf{r}) ~ \rev{\overline{H}_\textrm{in}}(\mathbf{r}-\mathbf{r}_\text{in}\al{,\rm}).
\label{Rrr_intermsofh_eq3}
\end{multline}
where we have assumed, in first approximation, that $\rev{\overline{H}_\textrm{in/out}}$ is constant over the frequency bandwidth. The occurrence of the sinc factor in the integrand of the last equation \rev{accounts for the time gating operation of \eqref{eq:sumRrr_freq}. Its spatial extent yields the expected axial resolution $\overline{\delta z}_0\sim c /2\Delta f$ in a broadband regime.}

\rev{Not surprisingly, the coherent sum of \eqref{eq:sumRrr_freq} drastically improves the axial resolution and contrast of the image $\overline{\mathcal{I}}(\mathbf{r},\Delta f)$  built from the diagonal\al{s} of $\Rrrbar(\Delta f)$ \eqref{eq:sumRrr_freq}}
\la{, revealing the micro-architecture of the calf tissues in the experiment (Fig.~\ref{fig:focusing2}(b$_1$)) and \rev{the anechoic region in the numerical simulations (Figs.~\ref{fig:focusing1}(b$_2$,b$_4$)}.} \alex{Fig.~\ref{fig:focusing2}(b$_2$) shows the \al{cross-section} $\overline{\mathbf{R}}_{xx}{(\Delta f,z)}$ of the \rev{experimental} FR matrix $\overline{\mathbf{R}}_{\mathbf{rr}}(\Delta f)$ at depth $z=18$ mm [dotted white line of 
\la{Fig.}~\ref{fig:focusing2}(\la{b$_1$})]. Compared to its monochromatic counterpart [Fig.~\ref{fig:focusing2}(a\la{$_2$})], the single scattering contribution along the diagonal of $\overline{\mathbf{R}}_{\mathbf{rr}}$ is enhanced \mat{with respect} to the off-diagonal coefficients. 
\la{This enhancement is due to the time-gating procedure; 
contributions from scatterers which sit above and below the focal plane have been eliminated, so the remaining singly-scattered echoes are located near the diagonal of $\overline{\mathbf{R}}_{\mathbf{rr}}$.}} 

\rev{Interestingly, numerical simulations show that the off-diagonal energy is now mainly due to multiple scattering events taking place 
\lc{at depths which are shallower than} the focal plane. While the monochromatic FR matrices did show a similar off-diagonal energy whatever the speckle level [see Figs.~\ref{fig:focusing1}(d$_1$,d$_3$)], the broadband one now exhibit\lc{s} a larger off-diagonal energy in the high speckle regime [see the comparison between Figs.~\ref{fig:focusing1}(d$_2$) and (d$_4$)]. This observation is quantitatively confirmed by investigating the broadband CMP intensity profile in Fig.~\ref{fig:focusing1}
\lc{(c)}. It now displays a confocal, steep peak mainly due to single scattering on top of a \rev{lower but wider} \rev{pedestal} linked to multiple scattering. The relative amplitude of the multiple scattering background with respect to single scattering is larger in the high speckle regime. In Sec.~\ref{sec6}, we will show how to map 
\lc{this relative amplitude quantitatively, and demonstrate its sensitivity to} speckle level.} 

\rev{Experimentally, the improvement in transverse resolution of the single scattering contribution can also be seen in} the broadband CMP intensity profile displayed in Fig.~\ref{fig:2Dcommon}(c)\la{, and in} 
its \al{cross-section} shown in Fig.~\ref{fig:focusing2}(b$_3$). 
\rev{Nevertheless}, although we are in a broadband regime, the 2D focal spot in Fig.~\ref{fig:2Dcommon}\mat{(c)} still exhibits a cigar-like shape. \rev{To illustrate why this is expected,}
\la{, we now express the ensemble average of the CMP intensity profile \eqref{paraxial} in the broadband regime under the paraxial approximation~\cite{lambert_phd}:}
\begin{equation}
\overline{I}_\textrm{av}(\rm,\Delta \mathbf{r}\al{,\Delta f}) = \mat{\frac{A}{\Delta f}} \al{\int_{f_-}^{f_+}} df  \left [ |\rev{\overline{H}_\textrm{in}}|^2  \al{\stackrel{\Delta x}{\circledast}} |\rev{\overline{H}_\textrm{out}}|^2 \right ] (\Delta \mathbf{r},\rm,f) \la{,} 
\label{IrDr_fromRrDr3}
\end{equation}
\la{where \mat{$A$} is a constant. Equation\ \eqref{IrDr_fromRrDr3} shows that,} \rev{unlike traditional speckle statistics~\cite{Wagner1983}, the CMP intensity profile essentially eliminates the phase components of the wave-field and yields a sum over the frequency bandwidth of the incoherent input-output PSF $|\rev{\overline{H}_\textrm{in}}|^2  \al{\stackrel{\Delta x}{\circledast}} |\rev{\overline{H}_\textrm{out}}|^2$}. This explains why 
\la{the axial resolution \rev{in Fig.~\ref{fig:2Dcommon}(c)} has not been improved by the time gating process, as 
\lc{was} \rev{the case} for the broadband FR matrix (Eq.~\ref{Rrr_intermsofh_eq3}).} Nevertheless, the CMP intensity evolution along $\Delta x$ 
still offers a way to estimate the transverse resolution of the imaging PSF in the broadband regime. Indeed, Fig.~\ref{fig:focusing2}(b3) 
shows that the transverse \rev{FWHM of $\overline{I}_\textrm{av}(\rm,\Delta \mathbf{r}\al{,\Delta f}) $ ($ \sim 0.5$ mm) remains \lc{far from} 
\rev{optimal} ($\delta x_0(f_c) \sim 0.25$ mm).} 

\la{In the next section, we aim to develop a better way to evaluate \rev{the aberration and multiple scattering levels}
in \rev{ultrasound imaging}. We will define quantitative parameter\rev{s} to measure the focusing quality and multiple scattering rate} at any pixel of the ultrasound image. \rev{The focusing criterion will correspond to the scaling factor that maps the transverse evolution of the CMP intensity profile, $I_\textrm{av}(\rm,\Delta \mathbf{r},\Delta f)$, onto its ideal diffraction-limited counterpart, $\bar{I}_0(\rm,\Delta \mathbf{r},\Delta f)$. A local multiple scattering rate will be also deduced from this fitting process.} 
To make our measurement quantitative, 
\la{the} theoretical prediction of \rev{$\bar{I}_0(\rm,\Delta \mathbf{r},\Delta f)$} should be as accurate as possible
\la{; in the following, we work towards this accuracy by developing a theoretical time-frequency analysis of the FR matrix.} 

\section{\label{sec5}Time-frequency analysis of the focused reflection matrix}
\alex{A time-frequency analysis of the \al{FR matrix} is required to investigate the evolution of absorption and scattering as a function of frequency. To do so, the coherent sum of the monochromatic FR matrices [Eq.~\ref{eq:sumRrr_freq}] can be performed over a smaller bandwidth $\delta f$ centered on a given frequency $f$: 
\begin{equation}
    \overline{\mathbf{R}}_\mathbf{rr} (f,\delta f) =   \willy{\frac{1}{\delta f}} \int^{f+\delta f}_{f-\delta f} d \la{f^\prime} ~ \Rrr(\la{f^\prime})
    \label{eq:sumRrr_freq2}
\end{equation}
\la{We have shown that} the axial dimension $\overline{\delta z}_0$ of the virtual transducers is inversely proportional to the frequency bandwidth $\delta f$. \la{Thus, a compromise must} 
be made between the spectral and axial resolutions. Here, the following choice has been made: $\delta f =0.5$ MHz and $\overline{\delta z}_0=3$ mm.}

\alex{Fig.~\ref{fig:focusing2}(c) shows the \rev{experimental} ultrasound images [Fig.~\ref{fig:focusing2}(c\la{$_1$})], FR matrices [Fig.~\ref{fig:focusing2}(c2)] and 
\la{CMP} profiles [Fig.~\ref{fig:focusing2}(c3)] 
\la{for} three different frequency bandwidths: 5-5.5 MHz, 7.25-7.75 MHz, 9.5-10 MHz. The axial resolution in each ultrasound image is of course deteriorated compared to the broadband image [Fig.~\ref{fig:focusing2}(b\la{$_1$})]
\la{; nevertheless,} the time-frequency analysis of the FR matrices yields the evolution of the SNR versus depth and frequency. At $z=18$~mm, for instance, the FR matrix at $f=9.75$~MHz exhibits a tiny confocal 
\la{intensity enhancement} on top of a predominant noise background (SNR$\sim$3dB). 
\la{Conversely,} the FR matrices at $f=5.25$ and $7.5$~MHz exhibit a CMP intensity profile 
\la{which more closely resembles} its broadband counterpart. This weak SNR at $9.75$~MHz can be partially explained by the finite bandwidth of the transducers ($5-10$~MHz)\rev{. Absorption and scattering} losses undergone by ultrasonic waves in soft tissues 
\la{also have} a strong impact on the ultrasound image. Fig.~\ref{fig:focusing2}(d) illustrates \rev{this effect} by displaying the spectrum of the confocal signal, $\left \langle \overline{\mathcal{I}}(\mathbf{r},f,\al{\delta f}) \right \rangle_x$, as a function of depth. This spectrum \la{shifts} towards low frequencies as a function of depth. This frequency shift \rev{seems} characteristic of absorption losses 
\la{in soft tissues}~\cite{duck1990acoustic}. \rev{However it is difficult to discriminate between absorption and scattering losses~\cite{Aubry2011} since the scattering mean free path $\ell_s$ can also vary across the frequency bandwidth.} }
\rev{Anyway, whether} 
due to absorption or scattering \rev{losses}, the 
variation of \la{the temporal frequency spectrum of} back-scattered echoes has a strong impact on the local resolution 
of the ultrasound images. In the next section, we show how to 
\la{incorporate this frequency dependence in the theoretical expression of the \rev{ideal CMP intensity profile}, in order to establish a \rev{precise and} quantitative focusing factor.}

\section{\label{sec6}The local focusing factor and multiple scattering rate} \label{par:focusing criterion}
\al{In this section, a local focusing criterion is established for the broadband ultrasound image. For \la{the} sake of lighter \la{notation}
, the dependence of each physical quantity with respect to $\Delta f$ is 
\la{omitted}.} Aberrations caused by 
\la{medium} heterogeneities degrade the resolution of the ultrasound image \alex{and induce a spreading of singly-scattered echoes over the off-diagonal coefficients of $\overline{\mathbf{R}}_\mathbf{rr}$}. In the speckle regime, it is difficult to 
\la{determine by eye whether} the image is aberrated, and if 
\la{so}, which areas are the most impacted. Interestingly, 
the CMP intensity profile can yield an unambiguous answer \la{to this question}. \alex{In the speckle regime, this profile 
yields the convolution between the incoherent input-output PSF averaged over the frequency bandwidth (Eq.\ \la{\ref{IrDr_fromRrDr3}}
). 
\la{While the incoherent input-output PSF is not \la{exactly} equal to the confocal imaging PSF (Eq.~\ref{Rrr_intermsofh_eq3}) it nevertheless} fully captures the impact of transverse aberrations. It thus constitutes a relevant observable for assessing 
\la{focusing} quality.} 

\rev{To measure a local focusing factor and multiple scattering rate, the normalized CMP intensity profile should be first decomposed as the sum of three contributions:
\begin{equation}
\label{CMP_decomp}
\begin{split}
{ \overline{I}_\textrm{av}(\rm,\Delta x)} & = \alpha_S(\rm)  \overline{I}_0 (\rm,F(\rm) \Delta x) \\  & +  \alpha_M(\rm)  \overline{I}_M (\rm,\Delta x) \\ &+ \alpha_N(\rm),
\end{split}
\end{equation}
\lc{where} $\alpha_S(\rm)$, $\alpha_M(\rm)$, and $\alpha_N(\rm)$, \lc{are the respective rates of} local single scattering, multiple scattering, and noise. 
These three quantities 
\lc{obey} the following relation: $\alpha_S(\rm)+\alpha_M(\rm)+\alpha_N(\rm)=1$. The first term accounts for the single scattering contribution and assumes that, 
\lc{on} average, aberrations 
reduce the effective numerical aperture by a focusing factor $F$. 
\lc{The spatial dependence of $F$} is thus a re-scaled version of the diffraction-limited CMP profile $\overline{I}_0 (\rm, \Delta x) $ that would be obtained in absence of aberrations. The second term in Eq.~\ref{CMP_decomp} accounts for the multiple scattering contribution. In first approximation, its spatial profile is assumed to be Gaussian~\cite{Aubry2007}, such that $\overline{I}_M (\rm,\Delta x)=\exp \left [-\Delta x^2/(2\sigma_M(\rm)^2 )\right ]$, with $\sigma_M$ the spatial extent of the diffuse halo. At last, the third term accounts for the noise background that is constant with respect to $\Delta x$ 
\lc{on} average.}

\rev{In the following, the free parameters $F$, $\alpha_S$, $\alpha_M$ and $\sigma_M$ will be determined through a fitting procedure\lc{, but we must} 
first accurately determine the reference CMP profile $\overline{I}_0 (\rm,\Delta x) $. }
To do so, the frequency spectrum of the ultrasound image \al{should be} taken into account. For each \al{CMP $\rm$, $\mathcal{I}(\rm,f)$} is an estimation of the desired frequency spectrum [Fig.~\ref{fig:focusing2}(d)]. 
\rev{$\overline{I}_0 (\rm,\Delta x)$ can then be computed by performing a frequency average of the theoretical CMI intensity profile, ${I}_0 (\rm,\Delta x,f)$~\eqref{I0}, weighted by the spectrum $\mathcal{I}(\rm,f)$ of the ultrasound image.}

\begin{figure*}
    \centering
    \includegraphics[width=14cm]{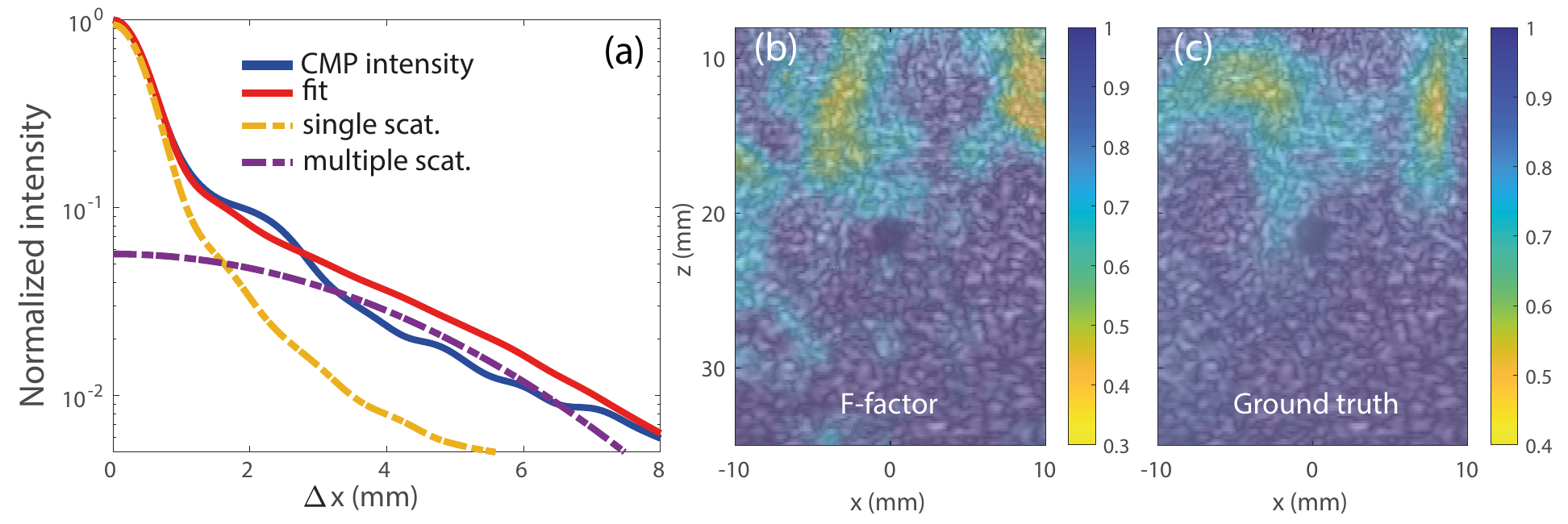}
    \caption{\rev{Numerical validation of the fitting process on the CMP intensity profile by Eq.~\ref{CMP_decomp}. (a) Result of the fit at $z=33$~mm in the low speckle regime: the fit parameters are $F=0.96$, $\alpha_M=5.5\%$ and $\sigma_M=3.4$~mm. (b) Map of the focusing factor $F(\rm)$ extracted from the CMP intensity profile averaged over spatial windows of size $L=4$~mm. (c) Ground-truth map of $F(\rm)$ computed from the measured PSFs $H_\textrm{in}$ and $H_\textrm{out}$ \eqref{IrDr_fromRrDr3}.}}
    \label{fig:fit}
\end{figure*}
\rev{Now that $\bar{I}_0 (\rm,\Delta x)$ is properly estimated, the CMP intensity profile is 
\lc{fit with} Eq.~\ref{CMP_decomp} using the \lc{Matlab} nonlinear programming solver fiminsearchcon. 
This fitting procedure is first validated by the numerical simulations. Fig.~\ref{fig:fit}(a) shows the result of this fit in the low speckle regime at an arbitrary depth $z=33$ mm. While the single scattering peak seems to be nicely fit by a rescaled version of $\overline{I}_0 $ ($|\Delta x|<2$ mm), a larger fit error is observed for the multiple scattering background ($|\Delta x|>2$ mm). Two main reasons can account for this discrepancy: (\textit{i}) the Gaussian model used to describe the transverse evolution of multiple scattering contribution that is only strictly valid in the diffusive limit~\cite{Aubry2007}
\lc{, and} (\textit{ii}) a lack of statistical average for smoothing the fluctuations induced by the random reflectivity of the medium. Indeed, a compromise has to be made between the size $L$ of the spatial window $W_{L}$ \eqref{eq:window} used to average the CMP intensity profile and the spatial resolution of the fitting parameters in \eqref{CMP_decomp}. This size $L$ should be small enough to 
\lc{preserve} the spatial variations of aberrations across the field-of-view, but large enough to smooth out the ultrasound speckle. Indeed, the speckle fluctuations in $\bar{I}_\text{av}(\rm,\Delta x)$ decrease as $1/N\in$ with $N\in \sim L^2/\delta x_0 \delta z_0 )$, the number of resolution cells contained in each spatial window. Here we chose \just{$L=3.5\lambda_c$} (with $\lambda_c$ the wavelength at the central frequency) for the numerical data which yields \just{$N\sim 45$}.}

Fig.~\ref{fig:fit}(b) shows the spatial evolution of the $F-$factor across the field-of-view. To evaluate the performance of the fitting procedure, the $F-$map is compared to its ground truth value (Fig.~\ref{fig:fit}(c)). This ground truth map is built using a local measure of the input and output PSFs, \revv{which is possible in numerical simulations. To do so, each PSF ${H}_\textrm{in/out}(\mathbf{r},\mathbf{r}_\textrm{in/out},t)$ is computed by simulating the propagation of an incident wave-field designed to focus at $\mathbf{r}_\textrm{in/out}$ from the plane wave (input) or transducer (output) basis. The PSF corresponds to the transmitted wave-field recorded in the focal plane with a time origin at the expected ballistic time. A temporal Fourier transform and a change a variable yields the corresponding de-scanned PSFs, ${H}_\textrm{in}^{(l)}$ and ${H}_\textrm{out}^{(l)}$, in the frequency domain. The ground truth CMP intensity profile is then provided by the following expression of $\overline{I}_\text{av}$~\cite{lambert_phd}:
\revv{
\begin{multline}
  \overline{I}_\text{av}(\rm,\Delta x ,\Delta f) =  \frac{B}{{\Delta f^2}} \int_{f_-}^{f_+} df \int_{f_-}^{f_+} df' \int d x \\ 
   {H}^{(l)}_\text{out}\left (x-{\Delta {x}}/{2},\rm,f \right)~ 
 {H}^{(l)*}_\text{out}\left (x-{\Delta x}/{2},\rm,f' \right) 
\\  
{ H}^{(l)}_\text{in}\left (x+{\Delta x}/{2},\rm ,f\right)~
 { H}^{(l)*}_\text{in}\left (x+{\Delta x}/{2},\rm,f ' \right) .
\end{multline}
where $B$ is a constant. This last expression is more general than \eqref{IrDr_fromRrDr3} since the frequency-dependence of the PSF is here considered.}}

\revv{Fig.~\ref{fig:fit} displays a satisfactory agreement between the $F-$factor and its ground truth value. Their slight discrepancy} can be explained by several reasons. First, as \lc{with} any ultrasound image, the $F-$map is shown as a function of an effective depth $z=c_0t/2$ that scales with the time-of-flight $t$ and the \rev{reference speed-of-sound $c_0$. On the contrary, the reference map displayed in Fig.~\ref{fig:fit}c evaluates the focusing factor at true depths in the medium. Second, the $F-$factor varies quite rapidly across the field-of-view for $z<20$ mm. This limited isoplanicity also contributes to the difference observed between the measured F-factor and its ground-truth value, 
\lc{which is especially evident} at shallow depths ($z<20$ mm). At larger depths, the slight discrepancy} \rev{observed around $x=-10$ mm between the measured F-}\rev{factor and its true value might be explained by 
\lc{slight imprecision at certain regions of the discrimination} between single and multiple scattering components. Again, the statistical fluctuations of the multiple scattering component and our hypothesis that the single scattering component can be described by a re-scaled version of $\overline{I}_0 (\rm,F(\rm) \Delta x)$ may explain, at least partially, the discrepancy observed between the measured $F-$map and its ground truth.}

\begin{figure}
    \centering
    \includegraphics[width=\linewidth]{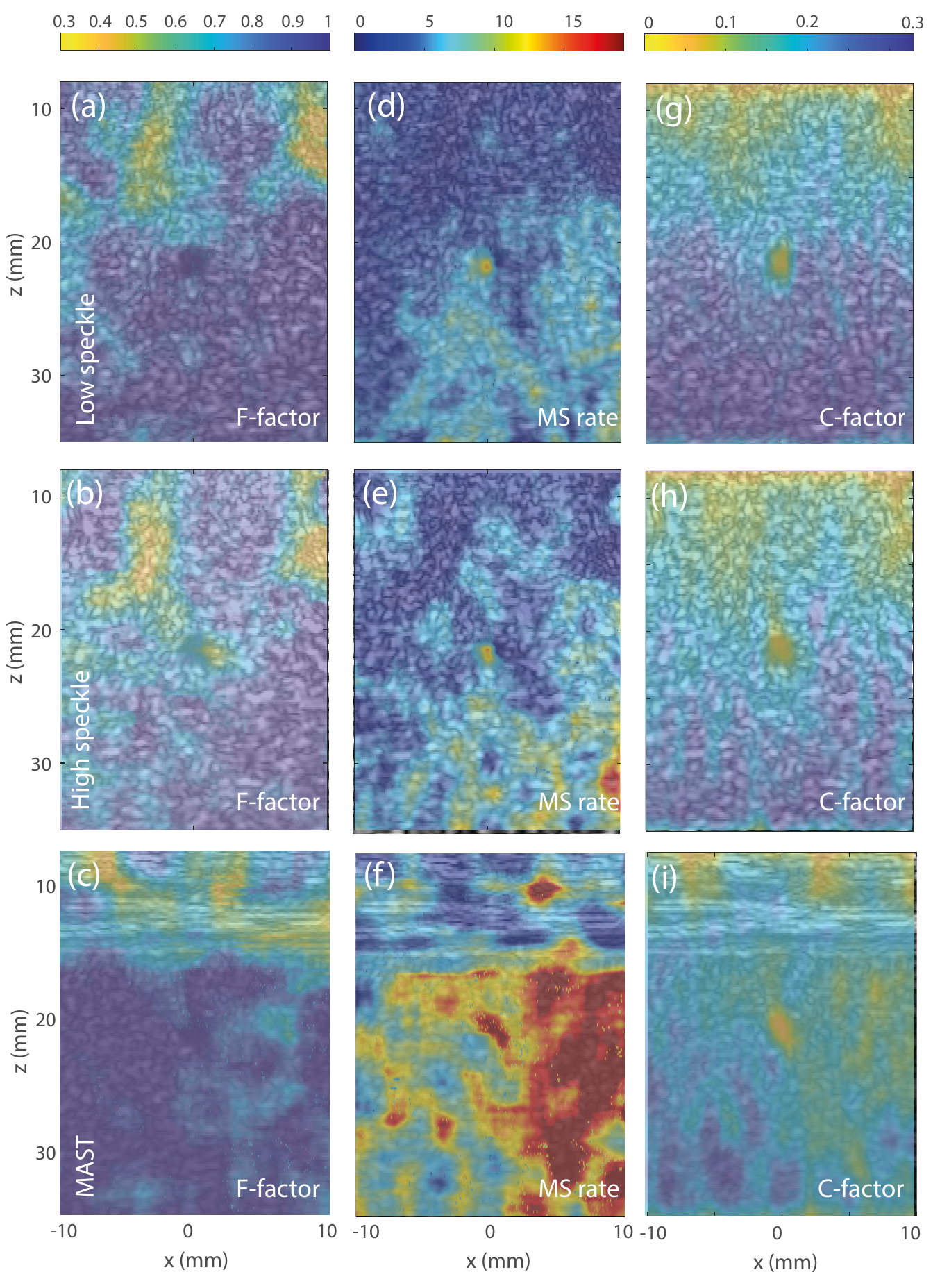}
    \caption{\revv{Mapping of local focusing quality and multiple scattering in numerical simulations. Maps of the (a,b,c) $F-$factor, (d,e,f) Multiple scattering rate and (g,h,i) $C-$factor in the low and high speckle regimes of the first simulation (Figs.~\ref{fig:rhoc}(a) and (b)) and in the low speckle regime of the second simulation (Figs.~\ref{fig:rhoc}(c) and (d)), respectively.}}
    \label{fig:numerics}
\end{figure}

\rev{
\rev{At shallow depths ($z<20 mm$),} the multiple scattering rate extracted by our fitting process shows a clear difference between the low (Fig.~\ref{fig:numerics}\revv{(c)}) and high (Fig.~\ref{fig:numerics}\revv{(d)}) speckle scattering regimes. On the contrary, the focusing factor $F$ remains remarkably constant in the same depth range [see the comparison between Figs.~\ref{fig:numerics}(a) and (b)]. This property highlights the fact that our approach enables 
\lc{discrimination} between 
aberration effects 
(quantified by $F$) 
\lc{and} multiple scattering phenomena (quantified by $\alpha_M$). 
Note that, at larger depths, the invariance of $F$ between the two scattering regimes \lc{(Eq.\ \ref{CMP_decomp})} is no longer 
\lc{strictly obeyed}. 
\lc{This effect (the difference between Figs.\ \ref{fig:numerics}(a) and (b) at $z>25$~mm)} may be explained by a different effective wave velocity of the medium in the high speckle regime.}

\revv{Beyond short-scale fluctuations of speed-of-sound and density in the medium, multiple scattering can also be caused by the strong impedance mismatch between tissues and connecting tissues in the abdominal wall. To illustrate this effect, a last numerical simulation considering the density and speed-of-sound distributions of Ref.~\cite{Mast1997} (Figs.~\ref{fig:rhoc}(c) and (d)) has been performed. The resulting maps of the focusing factor $F$ and multiple scattering rate $\alpha_M$ are displayed in Figs.~\ref{fig:numerics}(c) and (f). Compared with the initial simulation that considers smoother boundaries between tissues (Figs.~\ref{fig:numerics}(a) and (b)), $F$ and $\alpha_M$ are both higher in the speckle area behind the abdominal layer. This result can be understood as follows: Stronger impedance mismatch in the abdominal layer gives rise to more intense multiple reflection events, while smoothed variations of the speed-of-sound only deviate the trajectory of the incident and reflected waves, giving rise to more aberrations and less multiple scattering.}

\begin{figure*}
    \centering
    \includegraphics[width=\linewidth]{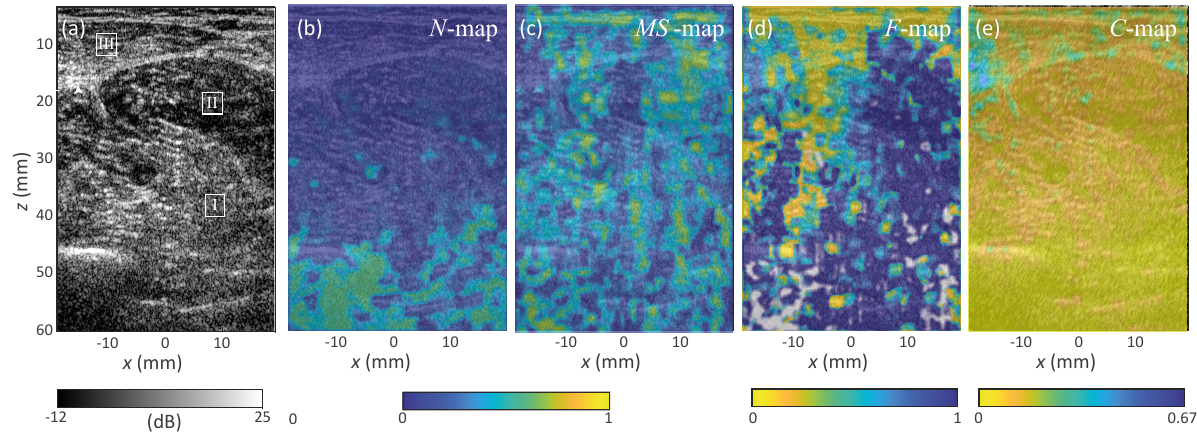}
    \caption{\rev{Mapping of local focusing quality and multiple scattering in the calf imaging experiment. (a) Ultrasound image of the calf. (b) Noise rate $\alpha_N (\rm)$. (c) Multiple scattering rate $\alpha_M (\rm)$. (d) $F$-factor. (d) $C-$factor. In (b,c) yellow areas correspond to a weak SNR and high multiple scattering rate, respectively. In (d,e) blue and yellow areas correspond to a high and low quality of focus respectively. In (d), gray areas highlights location where the estimation of the focusing criterion is not reliable because the incoherent background is too high ($\alpha_N+\alpha_M>0.75$)}} 
    \label{fig:US-image__focusingCriterion}
\end{figure*}

\rev{Fig.~\ref{fig:US-image__focusingCriterion} shows the application of the fitting process to the experimental data. \lc{In Fig.\ \ref{fig:US-image__focusingCriterion}(a), the original ultrasound image is shown. The extension $L$ of the spatial window $W_L$ has been set to} \just{1.5 mm}. \lc{Figs.\ \ref{fig:US-image__focusingCriterion}(b-d) show separate maps of the factors which affect the image quality.} Unlike in numerical simulations, electronic noise is present and a noise rate $\alpha_N$ should be quantified\lc{.} 
The evolution of $\alpha_N$ is 
\lc{shown in} Fig.~\ref{fig:US-image__focusingCriterion}(b). Unsurprisingly, a predominant noise occurs at large depth ($z>50$ mm) and in areas where the medium reflectivity is weak.} \rev{In any case, 
if a medical diagnosis is desired, the features of the ultrasound image 
in these areas \rev{should be more carefully interpreted. However, this is not the only 
\lc{factor which affects} the quality of the 
\lc{ultrasound image (Fig.~\ref{fig:US-image__focusingCriterion}(a))}. Clutter is also far from being negligible at shallow depths due to multiple reflection events between superficial layers and at larger depths because of multiple \rev{scattering processes in speckle (Fig.~\ref{fig:US-image__focusingCriterion}(c)). With regards to the single scattering contribution (Fig.~\ref{fig:US-image__focusingCriterion}(d)), the focusing quality shows strong variations across the field-of-view, particularly between the left and right parts of image. While high values} of $F$ (blue areas)  
{indicate good image reliability,} low values of $F$ (green/yellow areas) indicate a poor quality of focus. As expected, yellow areas seem to correspond to blurring of the ultrasound image [Fig. \ref{fig:US-image__focusingCriterion}(a)]. Gray areas correspond to the situation {in which} the estimation of the image resolution has failed because of a low SNR \rev{($\alpha_M+\alpha_N>0.75$)}. In these areas, the single scattering contribution is drowned out by a predominant incoherent background either caused by multiple scattering processes and/or electronic noise.}}

The ultrasound image shows different structures that are associated with their own speed of sound: (i) muscles tissues with three different fiber orientations \mat{[areas I, II, III on \la{F}\mat{ig.}~\ref{fig:US-image__focusingCriterion}(a)]}; (ii) two \la{v}eins located at $\mat{\lbrace x,z \rbrace} = {12,5}$~mm and ${-5,33}$~mm; \la{and} (iii) the fibula, located 
 \la{at} the bottom left of the figure ${-12, 45}$~mm. \rev{Fig.~\ref{fig:US-image__focusingCriterion}(d) reveals a poor focusing quality at shallow depth that \la{could} 
be explained by the near-field effects induced by the discrete sampling of the array. }
\rev{At larger depths, the $F-$ map exhibits a strong contrast between the right part of the image where the focusing factor $F$ is close to 1 and its left part where the focusing is degraded by the thicker and curved arrangement of superficial layers of skin, fat and muscle in area III.}

\revv{However, note there cannot be a direct correlation between the B-mode image and the $F-$ and $\alpha_M$-maps. Aberrations and multiple scattering are actually induced by wave velocity inhomogeneities and scatterers along the trajectories of the incident and reflected waves. For a given focusing point, the values of $F$ and  $\alpha_M$ are thus not related to the tissue architecture revealed by the B-mode image at this point but by speed-of-sound fluctuations and scattering properties of the medium above that point.} Retrieving the spatial distribution of the speed-of-sound \la{throughout the medium} from the quality of focus provided by the \revv{$F$-factor} is a difficult task that is beyond the scope of this paper.

\section{\label{sec7}Discussion}

The study presented here provides new insights into the construction of the FR matrix and \rev{a refined definition} of the focusing criterion \revv{and multiple scattering rate}.  Our results are representative of \textit{in-vivo} ultrasound imaging \la{in which} the medium under investigation\revv{, a human calf,} is 
composed \la{of} different kind\la{s} of tissues\la{, which can themselves be heterogeneous}. As the medium is composed of a mix of unresolved scatterers and specular reflectors, the scattering of ultrasound varies considerably in space, ranging from areas of strong to weak scattering. \revv{Moreover, ultrasound imaging of muscles is a relevant tool for monitoring neuro-muscular diseases~\cite{Wijntjes2020}. In that respect, a sharp quantification of aberrations and multiple scattering is crucial for mapping the speed-of-sound~\cite{Martiartu2022} and scattering anisotropy~\cite{Papadacci2014} in muscles. In shear wave elastography~\cite{Bercoff2004}, these phenomena can also have a detrimental impact on the inversion procedure that consists in extracting the muscle stiffness from ultrasound movies of shear wave propagation~\cite{Eby2013}.}

A direct and important application of this work would thus be to aid aberration correction by employing the parameter $F$ as a virtual guide star for adaptive focusing techniques. Currently, in the literature in this area, this role is performed by the coherence factor $C$~\cite{montaldo2011time,lediju2011short,dahl2017coherence}. \al{\la{Here, t}o highlight the benefit of our matrix approach with respect to the state-of-the-art, 
\la{we build a map of the standard coherence factor $C$ for the ultrasound image of the human calf. $C$}  
is equal to the ratio of the coherent intensity 
\la{to} the incoherent intensity of the realigned reflected wave-fronts recorded by the probe for each input focusing beam~\cite{robert2008green}. 
\la{Just as with the calculation of} the CMP intensity profile \eqref{average}, the raw coherence factor $C(\rin)$ is then spatially averaged over overlapping spatial windows to smooth the fluctuations due to the random reflectivity. The result is mapped onto the ultrasound image in Fig.~\ref{fig:US-image__focusingCriterion}\rev{e}. Compared to the $F-$map (Fig.~\ref{fig:US-image__focusingCriterion}\rev{d}), the coherence factor $C$ provides a weakly contrasted image of the focusing quality. To understand this difference, the analytical expression of $C$ can be derived under the isoplanatic hypothesis \eqref{isoplanatic}. While the CMP intensity enables a full retrieval of the spatial convolution between the incoherent input and output PSFs \eqref{IrDr_fromRrDr2b}, the coherence factor $C$ only probes this quantity at $\Delta x=0$~\rev{\cite{Mallart1994,Robert_thesis,robert2008green}}:
\begin{equation}
\label{Cth}
           C(\mathbf{r})= \left [ |\rev{\overline{H}_\textrm{in}}|^2 \stackrel{\al{\Delta x}}{\circledast}  |\rev{\overline{H}_\textrm{out}}|^2 \right] (\Delta x=0, \mathbf{r}).
\end{equation}
In the speckle regime, the coherence factor $C$ ranges from 0, for strong aberrations, to 2/3 in the ideal case~\cite{Mallart1994}. However, the presence of multiple scattering and noise can strongly hamper its measurement. As highlighted by \rev{Figs.~\ref{fig:US-image__focusingCriterion}{(b,c})}
, the incoherent input-output PSF, $|\rev{\overline{H}_\textrm{in}}|^2 \stackrel{\al{\Delta x}}{\circledast}  |\rev{\overline{H}_\textrm{out}}|^2$, 
\la{exhibits} an \rev{incoherent} background that is far from being negligible. Unlike the $C$-factor, the CMP intensity enables a clear discrimination between the single scattering contribution and the noise components. The $F-$factor 
\la{is sensitive only to aberrations,} since it is estimated after removing the incoherent background. 
This crucial feature accounts for 
\la{the difference in} behavior between $F$ and $C$ in Fig.~\ref{fig:US-image__focusingCriterion}, especially at large depths. \rev{While $F$ shows a 
\la{close to ideal} value of $1$ on the right part of the image beyond $z=10$ mm (Fig.~\ref{fig:US-image__focusingCriterion}(d)), $C$ exhibits a low value everywhere (Fig.~\ref{fig:US-image__focusingCriterion}(e)) because of a predominant incoherent background (Figs.~\ref{fig:US-image__focusingCriterion}(b) and (c)).} 
} \revv{This experimental observation is confirmed by the $C$-factor (Fig.~\ref{fig:numerics}(i)) obtained in the numerical simulation considering a strong impedance mismatch between tissues in the abdominal layer (Figs.~\ref{fig:rhoc}(c) and (d)). While the map of $F-$factor shows a relatively good focusing quality ($F>0.7$, see Fig.~\ref{fig:numerics}(c)) in the speckle area behind the abdominal layer, the $C-$factor exhibits a weak value ($C<0.2$, see Fig.~\ref{fig:numerics}(i)) due to a strong multiple scattering background ($\alpha_M>20 \%$, see Fig.~\ref{fig:numerics}(f)).} \la{This result} demonstrates one of the benefit\la{s} of \al{UMI} compared to standard ultrasound \la{imaging}: Probing the focusing quality in the focused basis \mat{drastically} improves \mat{the robustness to} multiple scattering and noise \mat{compared to} a direct cross-correlation of back-scattered echoes in the transducer basis.

\rev{Numerical simulations can highlight another advantage of UMI for a local estimation of the focusing quality. Figs.~\ref{fig:numerics}\revv{(g) and (h)} show the $C-$maps computed in the low and high speckle scattering regimes \revv{for the smoothed density and speed-of-sound distributions displayed in Figs.~\ref{fig:rhoc}(a) and (b)}. Contrary to the experiment \revv{and the numerical simulation mentioned above,} $C$ here shows a depth evolution in agreement with the ground truth focusing quality (Figs.~\ref{fig:numerics}(a) and(b)) because the multiple scattering background is weaker in this numerical simulation (Figs.~\ref{fig:numerics}\revv{(d) and (e)}). Nevertheless, the focusing factor $F$ grasps in a much more efficient way the transverse evolution of the focusing quality than the coherence factor $C$. This gain in terms of transverse resolution is provided by the input-output focusing operation inherent to UMI while the $C-$factor is estimated from a single focusing operation.}

\rev{Beyond the $C$-factor, the CMP intensity profile is an interesting analogue of the speckle autocorrelation (SAC) function, where the autocorrelation of the ultrasound image yields the \rev{pulse-echo PSF, $\bar{H}_{in} \times \bar{H}_{out}$, convolved against itself in the single scattering regime~\cite{Wagner1983}. Together, the CMP profile and SAC function seem to describe pieces of an even more general convolution over both $\Delta \mathbf{r}$ and $\Delta \rm$, with the CMP intensity profile being the case where $\Delta \mathbf{r} = \mathbf{0}$ and SAC function being the case where $\Delta \rm = \mathbf{0}$ (i.e. common input and output focusing points). Both are related measures of focusing quality in speckle: the CMP profile yields the proposed $F$ \rev{criterion, whereas the SAC FWHM is the traditional spatial resolution~\cite{Wagner1983}. Nevertheless, multiple scattering and/or electronic noise give rise to a peaked SAC function, potentially leading to the false impression of a diffraction-limited focus. 
\lc{In contrast,} those contributions manifest themselves as an \rev{incoherent background in the CMP intensity profile \lc{which 
allows them to be discriminated} from the single scattering component. The CMP approach is thus much more robust to multiple scattering and noise; moreover, it is not restricted to the speckle regime but can also be used to probe the focusing quality in presence of specular reflectors~\cite{lambert2020reflection}.}}}}

\al{In \mat{the second paper of the series}~\cite{lambert_ieee}, we present a matrix approach \rev{to} aberration correction \rev{for} \textit{in vivo} ultrasound imaging \rev{of a gallbladder.} An important aspect of that study is the use of $F$ to map the resolution of the image at each step of the aberration correction process. Again, numerical simulations are used \lc{to} validate the whole process and the measured $F$-factor is shown to be in agreement with its ground truth value.}

\section{\label{sec8}Conclusion}
\al{In summary, we successfully applied the concept of the FR matrix to the  \la{case of \textit{in vivo} ultrasound imaging of a human calf.} \rev{This approach has been priorly optimized and validated by means of numerical simulations.}
Thanks to the intuitive concept of virtual transducers, the FR matrix provides a wealth of information on the medium that goes well beyond a single confocal image. By splitting the location of the transmitted and received focal spots, the local resolution of the ultrasound image can be assessed at any pixel. By performing a time-frequency analysis of the reflection matrix, the contributions of single and multiple scattering and their impact o\la{n} the resolution and contrast 
\la{were} carefully investigated. This time-frequency study of the FR matrix paves the way towards a quantitative characterization of soft tissues by measuring parameters such as the attenuation coefficient or the scattering mean free path. In the accompanying paper~\cite{lambert_ieee}, the FR matrix will be used as a key building block of \al{UMI} for a local aberration correction. 
\la{Relatedly,} a focusing criterion 
\la{was} defined from the FR matrix in order to quantify 
the impact of aberrations {\rev{in the vicinity of each pixel of the ultrasound image}}. \al{Compared to the coherence factor generally used in the literature~\cite{Mallart1994,Hollman}, our focusing parameter is much more robust to noise and multiple scattering \rev{and \just{locally} maps the focusing quality with a much better transverse resolution.}} 
\la{Our} focusing parameter 
\la{is thus promising for use} in medical imaging as a reliability index of the ultrasound image. \just{Such an index would be of great use in the development and evaluation of advanced ultrasound imaging modes, and more importantly could help clinicians in their diagnostic. A quantitative and local assessment of image quality would indeed add strength to any ultrasound-based diagnostic, potentially avoiding any further and more invasive examination to the patient}. \just{Last but not least,  the focusing factor} can also be used as a guide star for adaptive focusing techniques\la{,} or \la{as} a local aberration 
\la{indicator} for \al{UMI}~\cite{lambert_ieee}. }

\section*{Acknowledgment}
\al{The authors wish to thank Victor Barolle, Amaury Badon and Thibaud Blondel whose own research works in optics and seismology inspired this study, \rev{as well as anonymous reviewers whose feedback \lc{has} helped us to improve the 
\lc{quality} of the manuscript.}}

\end{document}